\pdfoutput=1
\documentclass[prd,aps,preprint,amsmath,nofootinbib,amssymb,eqsecnum,showkeys,
superscriptaddress]{revtex4-1}

\usepackage{amsfonts}
\usepackage{multirow}
\usepackage{hhline}
\usepackage{caption}
\usepackage{subcaption}
\usepackage{amsmath}
\usepackage{slashed}
\usepackage{verbatim,graphics,graphicx,color,slashed}
\usepackage[usenames,dvipsnames]{xcolor}

\usepackage[normalem]{ulem} 

\usepackage{hyperref}

\newcommand{\tf}{\texorpdfstring}

\def \h1min{\Gamma_{h_1}^{\text{min}}}
\def \h2min{\Gamma_{h_2}^{\text{min}}}
\def \gev{~\text{GeV}}

\def \fbi{~\text{fb}^{-1}}

\begin{document}


\title{
Impact of the 125 GeV Higgs boson on the singlet fermion dark matter searches at the LHC
}


\author{P. Ko}
\email{pko@kias.re.kr}
\affiliation{School of Physics, Korea Institute for Advanced Study, 85 Hoegiro, Seoul 02455, Korea}

\author{Gang Li}
\email{gangli@phys.ntu.edu.tw}
\affiliation{Department of Physics, National Taiwan University, Taipei 10617, Taiwan}

\author{Jinmian Li}
\email{jmli@kias.re.kr}
\affiliation{School of Physics, Korea Institute for Advanced Study, 85 Hoegiro, Seoul 02455, Korea}

\date{\today}
\begin{abstract}
The search for singlet fermion dark matter at high-energy colliders is commonly analyzed with
a singlet scalar mediator, which however violates the standard model (SM) gauge invariance,
renormalizability and unitarity. These problems can be cured by introducing a mixing between the singlet
scalar $s$ and the SM Higgs boson $h$. Thus one has to consider two scalar mediators
$h_1$ and $h_2$, where $h_1$ is identified as the discovered 125 GeV Higgs boson. As a specific example, we consider the dark matter (DM) search in the $t \bar{t} + \slashed{E}_{T}$ channel.
According to the masses of dark matter and two scalar mediators, we classify the process into four cases.
By investigating the total cross sections and differential distributions, we find that the contribution
of the 125~GeV Higgs boson $h_1$ cannot be neglected in all cases and can even dominate
once $h_1$ is on-shell in dark matter production.
Further, we study the impact of $h_1$ on the LHC bounds of dark matter searches in the
hadronic and semileptonic channels with the integrated luminosity of $36~\text{fb}^{-1}$. Finally we make a brief comment that $h_1$ should be also considered in the vector DM search
at high-energy colliders.
\end{abstract}
\maketitle

\section{Introduction}
The framework of the simplified model~\cite{Buckley:2014fba,Haisch:2015ioa}(see also recent reports
~\cite{Abdallah:2015ter,Abercrombie:2015wmb}) has been widely used in studying the dark matter
(DM) phenomenology at colliders, where the interaction energy scale  can be much  higher than
the new physics scale so that the effective field theory approach is no longer valid
~\cite{Buchmueller:2013dya,Busoni:2013lha,Busoni:2014sya,Busoni:2014haa,Busoni:2014haa,
Pobbe:2017wrj}. In the  simplified model for singlet fermion DM, a singlet scalar $S$ is introduced, and its
interaction with the standard model (SM) quarks and the fermionic DM $\chi$ are assumed to be described by the following Lagrangian~\cite{Buckley:2014fba,Haisch:2015ioa}:
\begin{align}
\label{eq:simplified}
\mathcal{L}_{S}= - \sum_q g_q\dfrac{m_{q}}{v}\bar{q} q S - g_{\chi}\bar{\chi}\chi S,
\end{align}
where $q (=u,d,s,c,b,t)$ are the SM quarks and $v=246~\text{GeV}$.
The couplings $g_q$ and $g_\chi$ are usually chosen as
$g_{q}=g_{\chi}=1- 5$~\cite{Haisch:2015ioa,Arina:2016cqj} for simplicity and to guarantee a
sufficient DM production rate at colliders. Since the coupling of the mediator to a heavier flavor is
stronger, the DM production is dominated by the gluon fusion process with top quark involved at the Large Hadron Collider (LHC).
Then, DM can be searched through the mono-jet signature~\cite{Harris:2014hga,Harris:2015kda}
with the top quark in the loop or through the production in association with top quark(s).   Although the
process of DM in association with a top quark pair or single top quark has a small production cross
section, it can provide cleaner signals and more information on the  nature of DM and the interaction
form and has been studied extensively in the literature~\cite{Haisch:2015ioa,Arina:2016cqj,
Pinna:2017tay,Plehn:2017bys,Haisch:2016gry,Buckley:2015ctj,Dutta:2017sod}.
Especially, in Ref.~\cite{Arina:2016cqj} the $t\bar{t}+\slashed{E}_T$ signature is explored in the
simplified model in a comprehensive way and is compared to the sensitivity with mono-jet signature
as well as those without DM in the final state. In Refs.~\cite{Pinna:2017tay,Plehn:2017bys}, it is found
that the LHC search sensitivity can be significantly improved if one also includes DM production in
association with a single top quark.  In fact, recent experimental results at the LHC
~\cite{Sirunyan:2017hci,Aaboud:2017rzf,Sirunyan:2017leh,CMS:2018jww} already show that the
$t\bar{t}+\slashed{E}_T$ channel has a comparable sensitivity with the mono-jet channel.
Moreover, the shapes of angular separations between the two leptons from top quark decays are
found to be not only useful in resolving the coupling among the mediator and SM quark
~\cite{Haisch:2016gry,Haisch:2015ioa,Buckley:2015ctj} but also helpful in distinguishing DM
spins~\cite{Dutta:2017sod}.

However, one can easily find that the Lagrangian~\eqref{eq:simplified}  is not gauge invariant
under the SM gauge transformations,  since the SM left-handed quarks are in $SU(2)$ doublets
while right-handed ones are singlets.
In order that a singlet fermion DM can couple to SM particles in a renormalizable and gauge-invariant way, an economic way is to introduce a mixing between the singlet scalar $s$ with
the SM Higgs boson after the electroweak symmetry breaking.
As a result, there will be two physical scalar states ($h_1$ and $h_2$) mediating the SM and DM
interactions. These two scalars are identified as the 125 GeV Higgs boson and a second scalar
boson that can be either lighter or heavier than 125 GeV~\cite{Kim:2008pp,Baek:2011aa,Baek:2012uj},  respectively.  These two scalar bosons also
appear in the Higgs portal vector dark matter (VDM) model~\cite{Baek:2012se}.

The main phenomenological differences between the gauge-invariant Higgs portal DM models
and the simplified model~\eqref{eq:simplified} originating from the existence of the 125 GeV
Higgs boson has been discussed in Refs.~\cite{Baek:2015lna,Ko:2016ybp,Bauer:2016gys},
and references therein.  Especially, in Ref.~\cite{Ko:2016ybp} we have found that the interference
effect between two mediators can affect the LHC exclusion bounds considerably in some
parameter space, which was already reflected in Ref.~\cite{Baek:2015lna} to some extent.

In this work, we will investigate the impact of $h_1$ (with the mass of 125 GeV) in the singlet
fermionic DM (SFDM) model on the sensitivity of the LHC search for $t\bar{t} + \slashed{E}_T$.
Depending on the masses of DM and  mediators,  four different cases (named Case A,B,C and D) shall be  classified and discussed.
The collider bounds obtained from the simplified model framework are
applicable only in certain parameter space of Case C as we will show below.
In general, we will find that the simplified model cannot reproduce the results
derived from the renormalizable and gauge-invariant Higgs portal DM models.
Thus we conclude that  simplified models should be used for the DM search at colliders with
great caution, keeping in mind its limited success.

This paper is organized as follows: In Section~\ref{sec:SFDM}, we first introduce the Lagrangian describing
the renormalizable and gauge-invariant SFDM model and
compare it with the Lagrangian of the simplified model in Eq.~\eqref{eq:simplified}. Then, various constraints on the
SFDM model from the measurements of the 125~GeV Higgs boson as well as direct searches
for an extra Higgs boson are discussed.
In Section~\ref{sec:xsection}, we discuss the impact of the 125 GeV $h_1$  on the total cross
section as well as the differential cross section for $pp\to t\bar{t}\chi\bar{\chi}$
in four different cases. For Case A and
Case B, this process is dominated by the mediation of $h_1$ with $h_2 \sim S$
being irrelevant in most regions of $m_{h_2}$. For Case C and Case D, while the
cross section in the SFDM model can be larger or smaller than that in the simplified model, the
$p_{T}^{\chi\bar{\chi}}$ distribution in the SFDM model is always softer,  resulting in a smaller
cut efficiency. In Section~\ref{sec:limits}, we show the LHC bounds in the SFDM model and compare them with those in the simplified model.
In Section~\ref{sec:summary}, we summarize our results.

\section{SFDM model }
\label{sec:SFDM}
The renormalizable and gauge-invariant Lagrangian that describes the SM extended with a gauge
singlet fermion $\chi\sim (1,1,0)$\footnote{For $\chi$ in a nontrival representation of the SM gauge
groups, see Refs.~\cite{Chiang:2015fta,McKeen:2012av}} and a real singlet scalar $S$, i.e.,
the SFDM model, is~\cite{Kim:2008pp,Baek:2011aa,Chpoi:2013wga,Baek:2015lna,Ko:2016ybp,Esch:2013rta}
\begin{align}
\mathcal{L}=-(y_u\bar{Q}_L\tilde{H}u_R+y_d\bar{Q}_L H d_R)+\bar{\chi}(i\slashed{\partial}-
m_{\chi}-g_{\chi}S)\chi+\frac{1}{2}\partial_{\mu}S\partial^{\mu}S-V(H,S),
\end{align}
where $y_u$ and $y_d$ are the SM Yukawa couplings to the up- and down-type quarks, respectively, with suppressed
generation indices. The interaction $\bar{Q}_L\tilde{H}\chi_R$ is forbidden by the $U(1)_Y$ symmetry
while the interaction $\bar{L}_L\tilde{H}\chi_R$ can be discarded by a $Z_2$ symmetry under which
only $\chi$ is odd, {\it i.e.}, $\chi\to -\chi$~\cite{Ko:2016ybp} or by global $U(1)$  symmetry
~\cite{Fairbairn:2013uta}.

The scalar potential of the SFDM is given by \cite{Ko:2016ybp,Robens:2016xkb}
\begin{align}
\label{eq:potential}
V ( H, S) & = -\mu _H^2H^{\dagger}H + \lambda _H\left(H^{\dagger}H\right)^2
+\lambda _{HS}S^2 H^{\dagger}H+\mu _1 S
   H^{\dagger}H   \nonumber   \\
& +  \mu _0^3 S+\frac{1}{2} m_0^2 S^2+\frac{\mu _2 S^3}{3!}+\frac{\lambda _SS^4
   }{4!}.
\end{align}

The fields $h$ and $s$ are introduced after electroweak symmetric breaking as
\begin{equation}
H\to \left(
\begin{array}{c}
 0 \\
 \dfrac{h+v_H}{\sqrt{2}} \\
\end{array}
\right),\qquad S\to s+v_S,
\end{equation}
where $v_H$ and $v_S$ are the vacuum expectation values of $H$ and $S$, respectively.
The mass matrix of the scalar fields is
\begin{align}
\mathcal{M}^2
&=\left(
\begin{array}{cc}
 3 \lambda _H v_H^2-\mu _H^2+v_S^2 \lambda _{HS}+v_S \mu _1 & v_H \left(2 v_S
   \lambda _{HS}+\mu _1\right) \\
 v_H \left(2 v_S \lambda _{HS}+\mu _1\right) & \frac{1}{2} \left(2 m_0^2+2 v_H^2
   \lambda _{HS}+v_S^2 \lambda _S+2 v_S \mu _2\right) \\
\end{array}
\right),
\end{align}
so that the mass eigenstates $h_1$ and $h_2$ can be defined as
\begin{align}
\left(
\begin{array}{c}
 h_1 \\
 h_2 \\
\end{array}
\right) = \left(
\begin{array}{cc}
 \cos\theta &-\sin\theta \\
 \sin\theta &\cos\theta \\
\end{array}
\right) \left(
\begin{array}{c}
 h \\
 s \\
\end{array}
\right)
\end{align}
with
\begin{align}
\tan(2\theta)=\frac{4 v_H \left(2 \lambda _{HS} v_S+\mu _1\right)}{v_H^2 \left(2 \lambda
   _{HS}-6 \lambda _H\right)+2 \mu _H^2-2 \lambda _{HS} v_S^2+2
   m_0^2+\lambda _S v_S^2-2 \mu _1 v_S+2 \mu _2 v_S}.
\end{align}

It is well known~\cite{OConnell:2006rsp,No:2013wsa,OConnell:2006rsp,McKeen:2012av,
Chen:2014ask,Lewis:2017dme} that, for the most general
Lagrangian that describes a real singlet scalar extension of the SM, there is a \textit{shift symmetry}
$S\to S+\Delta_S$, which holds in the SFDM model even though $\chi$ is introduced
~\cite{Esch:2013rta}. Thus we can freely choose $\langle S \rangle=0$ without loss of generality
in this paper.

The minimal conditions
\begin{align}
\frac{\partial V}{\partial H}\bigg\rvert_{\langle H \rangle = v_H/\sqrt{2}}&=0,\\
\frac{\partial V}{\partial S}\bigg\rvert_{\langle S \rangle = 0}&=0
\end{align}
lead to
\begin{align}
\mu_H^2&=\lambda _H v_H^2,\\
\mu_0^3&=-\mu_1v_H^2/2.
\end{align}
In the basis of $\langle S \rangle =0$, the mixing between $S$ and $H$ comes solely from the term $\mu _1 S
   H^{\dagger}H $ in Eq.~\eqref{eq:potential}, and the mass matrix is simplified to
\begin{align}
\mathcal{M}^2=\left(
\begin{array}{cc}
 2 v_H^2 \lambda _H & v_H \mu _1 \\
 v_H \mu _1 & v_H^2 \lambda_{HS}+m_0^2 \\
\end{array}
\right).
\end{align}

Introducing the variable~\cite{OConnell:2006rsp}
\begin{align}
y\equiv \frac{-2\mu_{hs}^2}{\mu_{h}^2-\mu_{s}^2}
\label{eq:y}
\end{align}
with
\begin{align}
\mu_h^2=2 v_H^2 \lambda _H,\ \mu_s^2=v_H^2\lambda_{HS}+m_0^2,\
\mu_{hs}^2=v_H \mu _1,
\end{align}
the eigenvalues of the mass matrix can be expressed as
\begin{align}
m_{h_{1,2}}^2=\frac{\mu_{h}^2+\mu_{s}^2}{2}\pm \frac{\mu_{h}^2-\mu_{s}^2}{2}\sqrt{1+y^2},
\end{align}
where the sign $+ (-)$ corresponds to $m_{h_1}(m_{h_2})$, and the mixing angle
\begin{align}
\tan\theta=\frac{y}{1+\sqrt{1+y^2}},\quad \tan(2\theta)=y.
\end{align}
It is noted that $h_1$ is SM-like while $h_2$ is singlet-like for $\theta\in [-\pi/4,\pi/4]$.

In terms of mass eigenstates, the interaction Lagrangian of interest can be written as
\begin{align}
\mathcal{L}_{\text{int}} &
= - \left(h_1 \cos \theta + h_2 \sin \theta \right) \left( \sum_f \frac{m_f}{v_H} \bar{f} f
- \frac{2m^2_W}{v_H} W^+_\mu W^{-\mu} - \frac{m_Z^2}{v_H} Z_\mu Z^\mu  \right) \nonumber \\
     & + g_\chi \left(h_1 \sin \theta - h_2 \cos \theta \right) ~ \bar{\chi} \chi ~.~ \label{eq:lfdm}
\end{align}
The couplings of $h_1$ and $h_2$ to the SM fermion pair $(f\bar{f})$  or weak gauge boson pair
$(VV)$ with $V=W$ or $Z$ are given by
\begin{align}
\label{eq:coupling1}
g_{h_1xx}=c_{\theta}g_{hxx}^{\text{SM}},\quad g_{h_2xx}=s_{\theta}g_{hxx}^{\text{SM}},
\end{align}
where $xx=f\bar{f},VV$, $g_{hxx}^{\text{SM}}$ is the corresponding SM coupling,
$c_{\theta}\equiv \cos\theta$ and $s_{\theta}\equiv \sin\theta$.
The couplings of $h_1$ and $h_2$ to the DM pair $\chi\bar{\chi}$ are
\begin{align}
g_{h_1\chi\bar{\chi}}=s_{\theta}g_{\chi},\quad g_{h_2\chi\bar{\chi}}=c_{\theta}g_{\chi}.
\end{align}

Now we can ask if the usual simplified model, Eq.~\eqref{eq:simplified}, can be derived from the renormalizable and gauge-invariant model Lagrangian, Eq.~\eqref{eq:lfdm}.
For the phenomenology at $pp$ colliders,  there are two relevant energy scales in addition 
to the mass scales  ($m_{h_1}, m_{h_2}, m_\chi$, etc.)
in the Lagrangian: total center of mass energy $(\sqrt{s})$ and the center of mass energy at which the reaction actually occurs $(\sqrt{\hat{s}} \equiv x_1 x_2 s)$ with
$0\leq x_1 , x_2  \leq 1$ being energy fractions of partons inside the protons. $\sqrt{\hat{s}}$ is relevant since it is nothing but the characteristic 
scale of the hard scattering at parton levels, whereas $\sqrt{s}$ is important since it is highest 
energy scale provided by  the $pp$ colliders.

Above all, let us notice that the model Lagrangian considered in Ref.~\cite{Arina:2016cqj}
can be obtained by simply removing  by hand (or integrating out) the $h_1$ field in Eq. (\ref{eq:lfdm}).
However it is clear  that  this procedure is justified only if all the quantities $\sqrt{s}$, $\sqrt{\hat{s}}$, and
$m_{h_2}$ are (much) smaller than  $ m_{h_1} = 125$ GeV or $\sqrt{\hat{s}}$ is resonantly enhanced at $m_{h_2}$.
Otherwise, we cannot ignore (or integrate out) $h_1$
in the model Lagrangian.   One has to  include the effects of both $h_1$ and $h_2$,  since the
interference between them could be important in certain cases
\cite{Baek:2015lna,Ko:2016xwd,Ko:2016ybp,Kamon:2017yfx,Dutta:2017sod}.
There is no systematic way to derive the usual SFDM model, Eq.~\eqref{eq:simplified}, as a proper effective field
theory from Eq. (\ref{eq:lfdm}), if $m_{h_2} > m_{h_1}$. In the following, we will show explicitly
a number of examples where the role of $h_1$ is significant and the results are qualitatively different
from those in Ref.~\cite{Arina:2016cqj}.

Due to the mixing of $s$ and $h$ as in the SFDM model, the triple scalar couplings of $h_1-h_2-h_2$ and $h_2-h_1-h_1$ are also relevant, which are given by~\cite{Chpoi:2013wga,No:2013wsa,
Chen:2014ask,Dupuis:2016fda,Huang:2017jws,Falkowski:2015iwa,Lewis:2017dme}
\begin{align}
\label{eq:lam1-lam2}
\lambda_{122}&=3\lambda_H v_H s_{\theta}^2c_{\theta}-2\lambda_{HS} v_H c_{\theta}(s_{\theta}^2
-c_{\theta}^2/2)-\frac{1}{2}\mu_1s_{\theta}(s_{\theta}^2-2c_{\theta}^2)-\frac{1}{2}\mu_2c_{\theta}
^2s_{\theta},\\
\lambda_{211}&=3\lambda_H v_H c_{\theta}^2s_{\theta}-2\lambda_{HS} v_H s_{\theta}(c_{\theta}^2
-s_{\theta}^2/2)+\frac{1}{2}\mu_1c_{\theta}(c_{\theta}^2-2s_{\theta}^2)+\frac{1}{2}\mu_2s_{\theta}
^2c_{\theta},
\end{align}
respectively.

In the SFDM model with $v_S=0$, the free parameters were chosen to be $m_{h_1}$, $m_{h_2}$, $\theta$, $\lambda_{HS}$, $\mu_2$, $\lambda_S$, $v_H$, $m_{\chi}$ and $g_{\chi}$~\cite{Esch:2013rta}. Identifying $m_{h_{1}}$ and $v_H$ as $m_{h_{1}}=125~\text{GeV}$, $v_H=246~\text{GeV}$, we however choose the following parameters for the convenience of collider phenomenology
\begin{align}
m_{h_{2}},\theta,\lambda_{1},\lambda_{2},m_{\chi},g_{\chi},
\end{align}
where $\lambda_{1}$ and $\lambda_{2}$ are normalized triple scalar couplings defined as
\begin{align}
\label{eq:lam1_lam2}
\lambda_{1}\equiv\frac{\lambda_{122}}{\lambda_{\text{SM}}},
\quad\lambda_{2}\equiv\frac{\lambda_{211}}{\lambda_{\text{SM}}},\quad \text{with}\quad
\lambda_{\text{SM}}=\frac{m_{h_1}^2}{2v_H}.
\end{align}

The production cross sections of $h_1$ and $h_2$ at the LHC can be expressed as
\begin{align}
\sigma(p p \to h_1+X)&=c_{\theta}^2\sigma^{\text{SM}}(p p \to h(m_{h_1})+X),\\
\sigma(p p \to h_2+X)&=s_{\theta}^2\sigma^{\text{SM}}(p p \to h(m_{h_2})+X).
\end{align}

The total widths of $h_1$ and $h_2$ are\footnote{We have neglected the widths of three-body
decays since their contributions are usually subdominant~\cite{Dupuis:2016fda}.}
\begin{align}
\Gamma_{h_1} &=c_{\theta}^2\Gamma^{\text{SM}}_{h}(m_{h_1})+\Gamma(h_1\to h_2h_2)
+\Gamma(h_1\to \chi\bar{\chi}),\\
\label{eq:width_h2}
\Gamma_{h_2} &=s_{\theta}^2\Gamma^{\text{SM}}_{h}(m_{h_2})+\Gamma(h_2\to h_1h_1)
+\Gamma(h_2\to \chi\bar{\chi}),
\end{align}
where $\Gamma^{\text{SM}}_{h}(m_{h_1})$ and $\Gamma^{\text{SM}}_{h}(m_{h_1})$ correspond
to the total decay width of the SM Higgs boson~\cite{Heinemeyer:2013tqa} with the mass being
$m_{h_1}$ and $m_{h_2}$, respectively. The decay widths of $h_1$ and $h_2$ into $\chi\bar{\chi}$ are
\begin{align}
\Gamma(h_1\to \chi\bar{\chi})&=\dfrac{s_{\theta}^2g_{\chi}^2m_{h_1}}{8\pi }(1-4m_{\chi}^2/
m_{h_1}^2)^{3/2}\theta(m_{h_1}-2m_{\chi}),\\
\Gamma(h_2\to \chi\bar{\chi})&=\dfrac{c_{\theta}^2g_{\chi}^2m_{h_2}}{8\pi }(1-4m_{\chi}^2/
m_{h_2}^2)^{3/2}\theta(m_{h_2}-2m_{\chi}),
\end{align}
where the Heaviside step function $\theta(x)=1$ for $x >0$ and $\theta(x)=0$ for $x \le 0$ and
$m_{\chi}$ is the DM mass. The widths of $h_1$ and $h_2$ into $h_2h_2$ and $h_1h_1$ are
~\cite{No:2013wsa,Huang:2017jws,Chen:2014ask,Lewis:2017dme,Falkowski:2015iwa}
\begin{align}
\Gamma(h_2\to h_1h_1)&=\dfrac{\lambda_{2}^2\lambda_{\text{SM}}^2\sqrt{1-4m_{h_1}^2/
m_{h_2}^2}}{8\pi m_{h_2}}\theta(m_{h_2}-2m_{h_1}),\\
\Gamma(h_1\to h_2h_2)&=\dfrac{\lambda_{1}^2\lambda_{\text{SM}}^2\sqrt{1-4m_{h_2}^2/
m_{h_1}^2}}{8\pi m_{h_1}}\theta(m_{h_1}-2m_{h_2}),
\end{align}
respectively. It is clear that the above widths are not sensitive to the signs of $\lambda_1$ and 
$\lambda_2$. Therefore we shall concentrate on the magnitudes of $\lambda_1$ and $\lambda_2$ 
in the collider study for the SFDM model.

Since $h_2$ may also decay into an extended dark sector other than $\chi\bar{\chi}$, similar to the
simplified models~\cite{Buckley:2014fba,Haisch:2015ioa,Abdallah:2015ter,
Abercrombie:2015wmb} we can introduce the \textit{minimal} total width of $h_2$ in the SFDM
model~\cite{Baek:2015lna,Ko:2016ybp}:
\begin{align}
\h2min = \Gamma_{h_2}~\text{with}~\Gamma(h_2\to h_1h_1)=0.
\end{align}
It should be emphasized that the minimal total width of $h_2$ in the SFDM model also includes the
partial decay width into $WW^*$ and $ZZ^*$, without which as in the simplified model the cancellation~\cite{Maltoni:2001hu} between diagrams with $h_2WW$ and $h_2t\bar{t}$ interactions in DM
production in association with single top quark $pp\to tj\chi\bar{\chi}$ does not occur and the
sensitivity in $pp\to tj\chi\bar{\chi}$ can be even comparable to that in $pp\to t\bar{t}\chi\bar{\chi}$~\cite{Pinna:2017tay,Plehn:2017bys}.

The mixing angle ${\theta}$ is constrained by the Higgs signal strength measurements:
$\sin^2{\theta}\lesssim0.12$ at 95\% confidence level (C.L.)~\cite{Aad:2015pla,
Khachatryan:2016vau,Robens:2016xkb}, while constraints from heavy Higgs boson direct
searches~\cite{Khachatryan:2015cwa,Aad:2015kna,ATLAS:2017spa,Lewis:2017dme} and the
electroweak precision observables~\cite{Robens:2016xkb,Falkowski:2015iwa,
Huang:2017jws,Baek:2011aa} are found to be weaker than the Higgs signal strength
measurements.~\footnote{It is however found  in Refs.~\cite{Robens:2016xkb,Huang:2017jws}
that the constraint on $\theta$ from direct searches can be slightly stronger for the heavy Higgs
boson mass below 450 GeV in the SM with a singlet scalar. But it is relaxed in the SFDM model
if $m_{\chi}<m_{h_2}/2$ so that $h_2 \rightarrow \chi \bar{\chi}$ is kinematically open.}
In this paper, we will fix $\sin\theta$ to be $0.2$ conservatively.

\begin{figure}[!thb]
\includegraphics[width=0.38\textwidth]{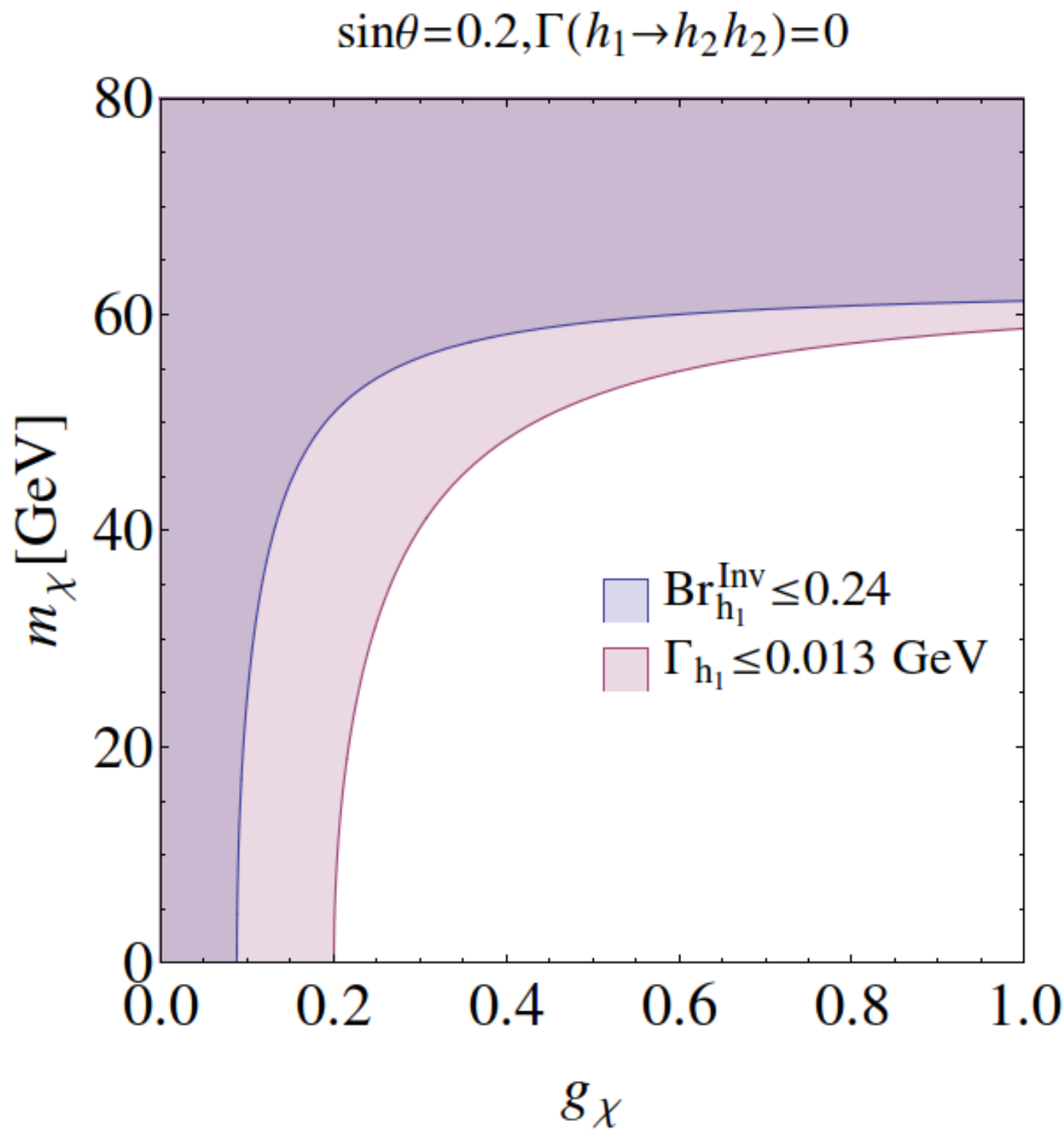}
\includegraphics[width=0.4\textwidth]{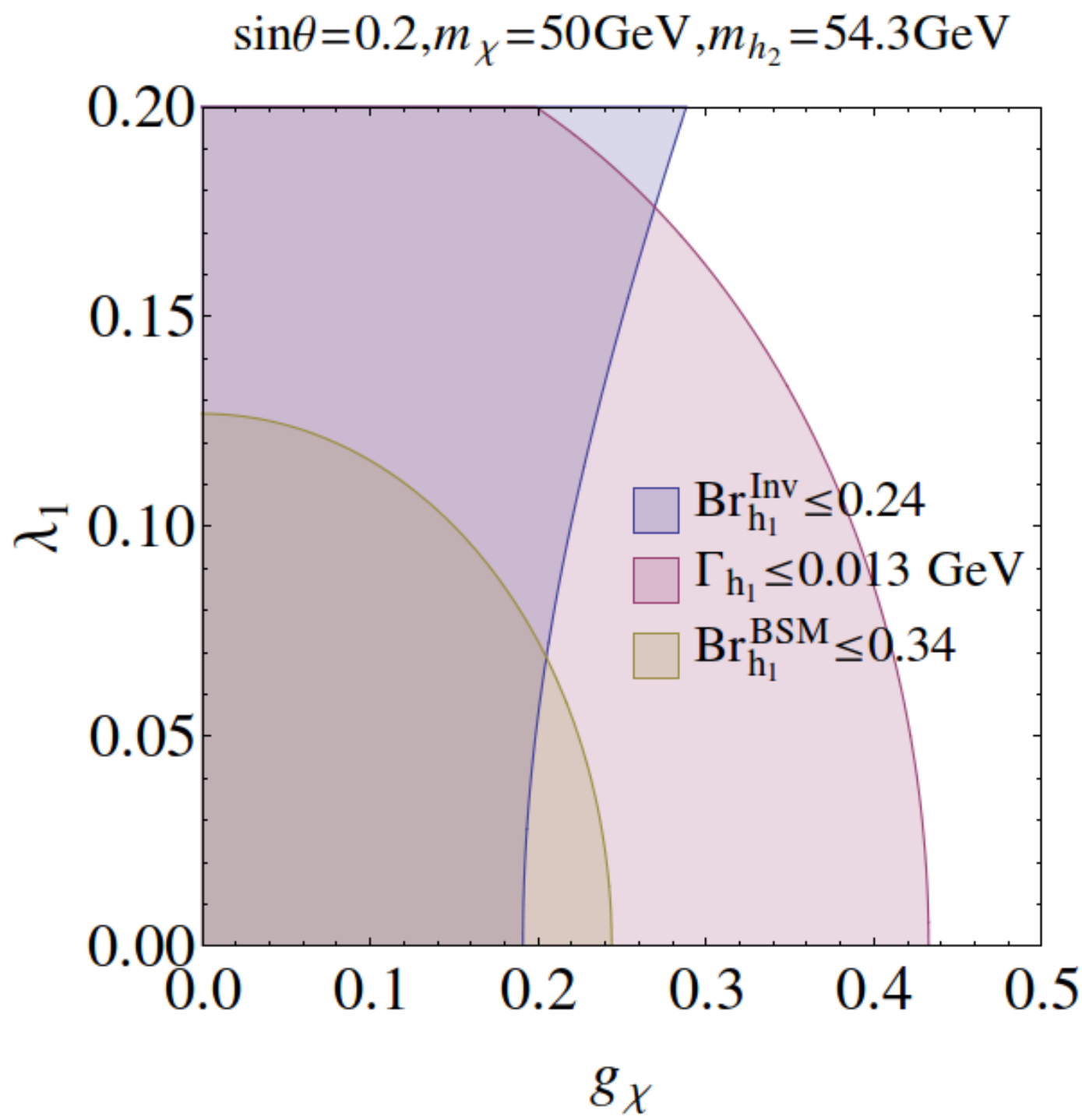}
\caption{The allowed regions by the invisible decay branching ratio (blue), total width (red) and BSM decay branching ratio (orange) of $h_1$ in the SFDM model with $\sin\theta=0.2$. Left panel: $\Gamma(h_1\to h_2h_2)=0$. Right panel: $m_{\chi}=50\gev$ and $m_{h_2}=54.3\gev$.}
\label{fig:invisible_total_h_1}
\end{figure}

The current 95\% C.L. upper limits on the invisible decay branching ratio and
the total width of the 125~GeV Higgs boson are 0.24
~\cite{Khachatryan:2016whc,Aad:2015pla,CMS:2018awd} and 0.13 GeV~\cite{Khachatryan:2016ctc}, respectively.
In the left panel of Fig.~\ref{fig:invisible_total_h_1}, we show the constraints in the $g_{\chi}-m_{\chi}$ plane with
$\sin\theta=0.2$ and $\Gamma(h_1\to h_2h_2)=0$, which indicates that the constraint from the
invisible decay branching ratio of the 125 GeV Higgs boson is stronger than that from its total width. If $m_{h_1}>2m_{h_2}$, the decay channel $h_1\to h_2h_2$ is kinematically allowed. In the right panel of Fig.~\ref{fig:invisible_total_h_1}, the constraints in the $g_{\chi}-\lambda_1$ plane from  the invisible decay branching ratio and total width of $h_1$ and the branching ratio into the beyond SM (BSM) decays $\text{Br}^{\text{BSM}}_{h_1}<0.34$~\cite{Khachatryan:2016vau}~\footnote{We  find that light boson direct
searches~\cite{Khachatryan:2017mnf} 
can also directly constrain the triple scalar coupling $\lambda_1$. But as found in Ref~\cite{Baek:2017vzd}, the constraint from the light boson direct searches is much weaker than that from the BSM decay branching ratio. } with $\sin\theta=0.2$, $m_{\chi}=50\gev$ and $m_{h_2}=54.3\gev$ are displayed. We will show in Section \ref{sec:xsection} that, such benchmark values of $m_{\chi}$ and $m_{h_2}$ are appropriate for Case B.

On the other hand, searches for di-Higgs production play a key role in the determination of the triple scalar coupling $\lambda_{2}$.
The cross section  of $pp\to h_1h_1$ can be parameterized as~\footnote{In reality, the coupling of
$h_1-h_1-h_1$ can contribute to the non-resonant production of $h_1h_1$, which is neglected here.}
\begin{align}
\sigma(pp\to h_1h_1)&=\sigma^{\text{SM}}(pp\to h_2)\times s_{\theta}^2\dfrac{\lambda_2^2
f(m_{h_2})}{s_{\theta}^2+c_{\theta}^2g_{\chi}^2g(m_{h_2})+\lambda_2^2f(m_{h_2})},
\end{align}
with
\begin{align}
f(m_{h_2})&=\dfrac{\lambda_{\text{SM}}^2\sqrt{1-4m_{h_1}^2/m_{h_2}^2}}{8\pi
m_{h_2}\Gamma^{\text{SM}}(m_{h_2})},\\
g(m_{h_2})&=\dfrac{m_{h_2}(1-4m_{\chi}^2/m_{h_2})^{3/2}}{8\pi \Gamma^{\text{SM}}(m_{h_2})}
\theta(m_{h_2}-2m_{\chi}).
\end{align}

\begin{figure}[!thb]
\includegraphics[width=0.4\textwidth]{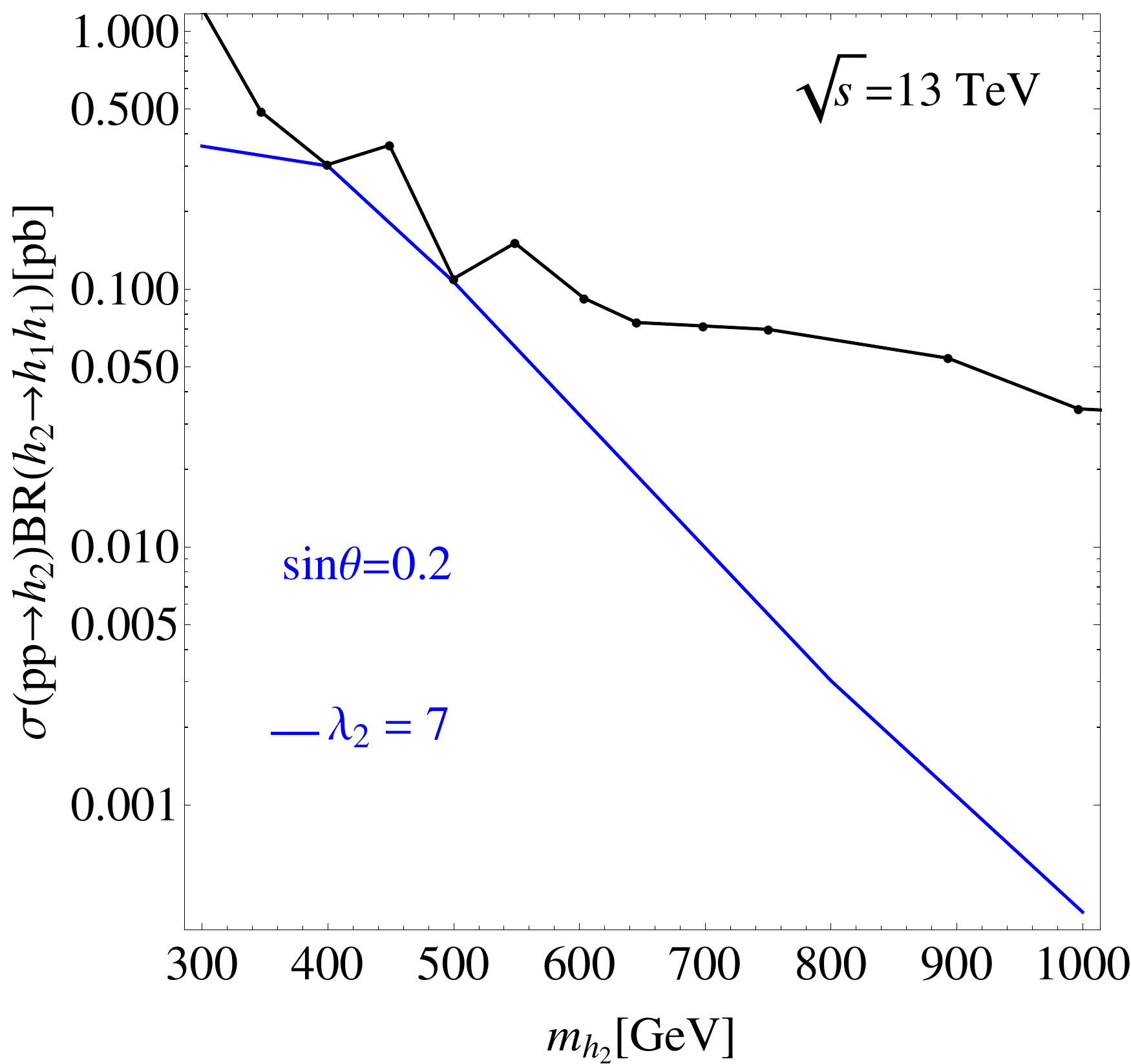}
\caption{Constraints on $\lambda_{2}$ from resonant di-Higgs searches at the 13 TeV
LHC. }
\label{fig:constraint_on_lambda2}
\end{figure}

Figure~\ref{fig:constraint_on_lambda2} shows the di-Higgs production cross section in the SFDM model (assuming $\Gamma(h_2\to \chi\bar{\chi})=0$ and
$\sin\theta=0.2$) alongside the combined upper limit at the 13 TeV LHC~\cite{CMS:2018obr}, which implies that $\lambda_2<7$. On the other hand, the theoretical constraint on $\lambda_2$ 
can be found in Ref.~\cite{Lewis:2017dme}, which is $\lambda_2 \lesssim 10$.  
For $\Gamma(h_2\to \chi\bar{\chi})\neq 0$,  larger $\lambda_2$ could be allowed depending 
on the mass $m_{\chi}$ and also the coupling $g_{\chi}$, which will not be studied in details in this paper.

\begin{figure}[!thb]
\includegraphics[width=0.4\textwidth]
{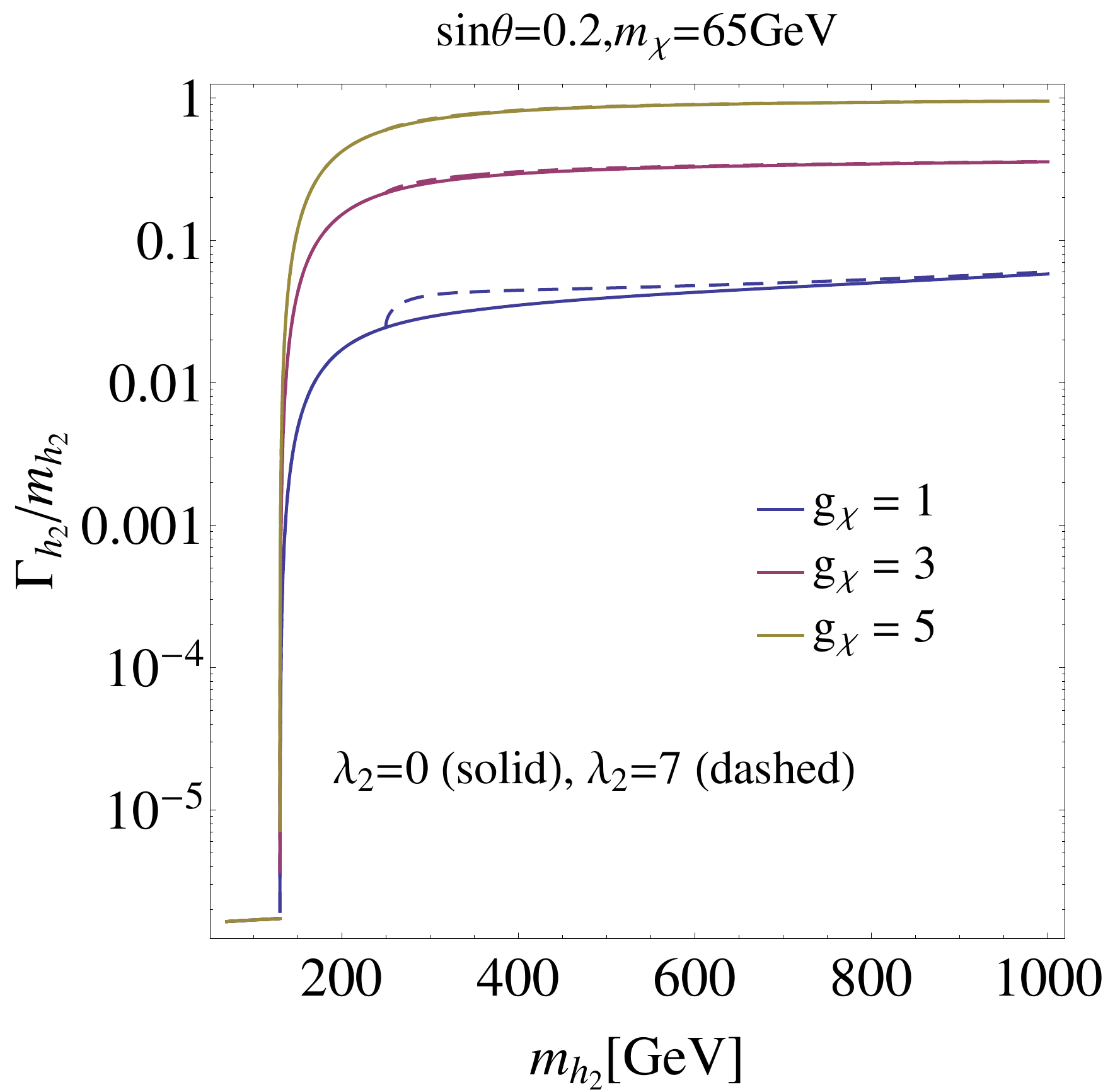}
\caption{The ratios $\Gamma_{h_2}/m_{h_2}$ with respect to $m_{h_2}$ for different $g_{\chi}$
values, where $\sin\theta=0.2$, $m_{\chi}=65~\text{GeV}$. For larger $m_{\chi}$, the ratio is smaller.
The ratios with (without) including $h_2\to h_1h_1$ partial width are denoted by solid (dashed) curves.}
\label{fig:width_h2}
\end{figure}

\begin{figure}[!thb]
\includegraphics[width=0.40\textwidth]
{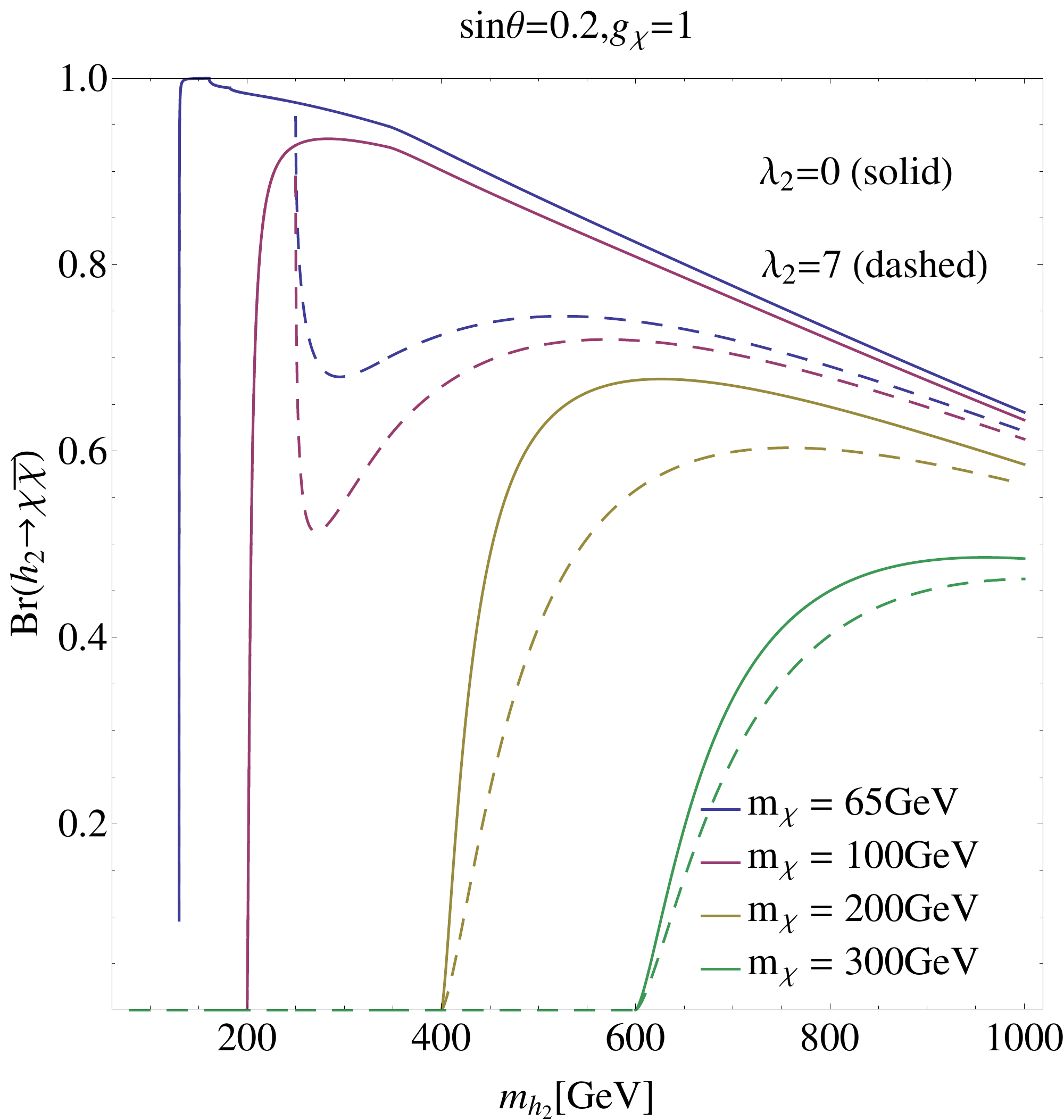}
\includegraphics[width=0.40\textwidth]
{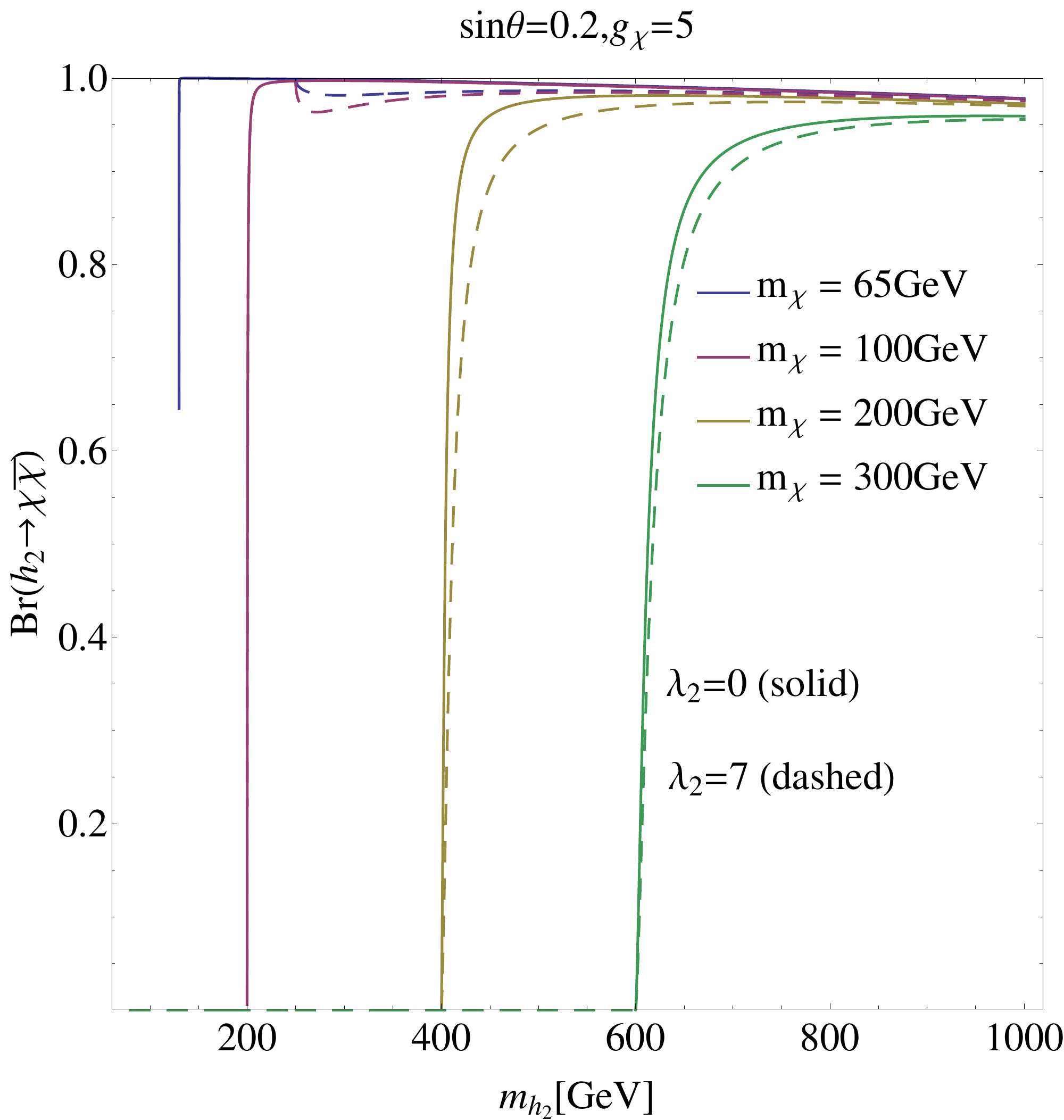}
\caption{Branching ratios of $h_2\to \chi\bar{\chi}$ with varying $m_{\chi}$, where $\Gamma_{h_2}$
is defined in Eq.~\eqref{eq:width_h2}. Left: $g_{\chi}=1$; right: $g_{\chi}=5$. The branching ratios
without (with) the decay $h_2\to h_1h_1$ are denoted by solid (dashed) curves.}
\label{fig:branching_ratios}
\end{figure}

In Fig.~\ref{fig:width_h2}, the total width of $h_2$ for $\sin\theta=0.2$, $m_{\chi}=65~\text{GeV}$ (for a larger $m_{\chi}$, the total width is smaller)
is displayed. We can find that for $g_{\chi}\lesssim 1$, the ratio $\Gamma_{h_2}/m_{h_2}$ is below
$10\%$ and $\Gamma_{h_2}/m_{h_2}<1$ can still be satisfied even with $g_{\chi}=5$. To evaluate
the impact of the decay $h_2\to h_1h_1$, we further show the branching ratio of $h_2\to
\chi\bar{\chi}$ for various parameter choices in Fig.~\ref{fig:branching_ratios}. One can find that
including $h_2\to h_1h_1$ can decrease $\text{Br}(h_2\to \chi\bar{\chi})$ especially for smaller
$g_{\chi}$. However it does not affect much the behavior of interplay between $h_1$ and $h_2$
in the DM search in Section~\ref{sec:xsection}, so we will keep $\lambda_2=0$ for simplicity hereafter.

In the SFDM model, the DM pair can annihilate into either SM gauge bosons/fermions through $h_1/h_2$ mediation or scalar bosons through $t$-channel/$s$-channel process if it is kinematically allowed. The annihilation cross section of the former is proportional to $g_\chi^2$ while that of the latter is proportional to $g_\chi^4$ ($g^2_\chi$) for $t$-channel ($s$-channel) annihilation. So a large $g_\chi$ is required to annihilate the DM effectively, which renders the DM direct detections quite stringent (except in the resonant region of the $s$-channel process $m_{\chi} \sim m_{h_{1,2}}/2$). The details of the DM thermal relic density 
and the spin-independent DM-nucleon scattering cross section for benchmark points in each case are provided in the Appendix. 
In order to guarantee that the DM has a relic density below the observation and keep consistent with the DM direct detections, the SFDM model should be generalized beyond the minimal setup of the Higgs portal SFDM model. 
New DM annihilation channels should be introduced, such as $\chi \bar{\chi} \to Z' Z'$ if DM is charged under a dark $U(1)$ gauge group~\cite{Brahm:1989jh,Ko:2014nha,Ko:2014uka,Baek:2014kna}, or coannihilation channels if there are more particles in the DM sector that have a mass close to the DM~\cite{Griest:1990kh}. Then the dark matter direct detection constraints can be weakened or even completely evaded. 
Moreover, the particle $\chi$ discussed in the current paper may correspond to a heavier dark state in the dark 
sector that can decay into the genuine DM candidate. Then, as long as the heavier dark state(s) does not leave any signal at the detector (due to a long lifetime or invisible decay), it will produce the same collider phenomenology as the SFDM model \footnote{Work in progress.}.

\section{\tf{$t\bar{t}+\slashed{E}_T$}{ttbar+MET} signature}
\label{sec:xsection}

Now  we are ready to study the impact of the $125\gev$ Higgs boson $h_1$ in the DM search with
$t\bar{t}+\slashed{E}_T$ signature. Previous studies in the mono-jet, VBF and mono-V signatures
can be found in Refs.~\cite{Baek:2015lna,Ko:2016ybp}.  The $t\bar{t} + \slashed{E}_T$ channel
at the 8 TeV LHC was preformed in Ref.~\cite{Baek:2015lna}, and compared with the CMS results.

\begin{figure}[!thb]
\includegraphics[width=0.3\textwidth]{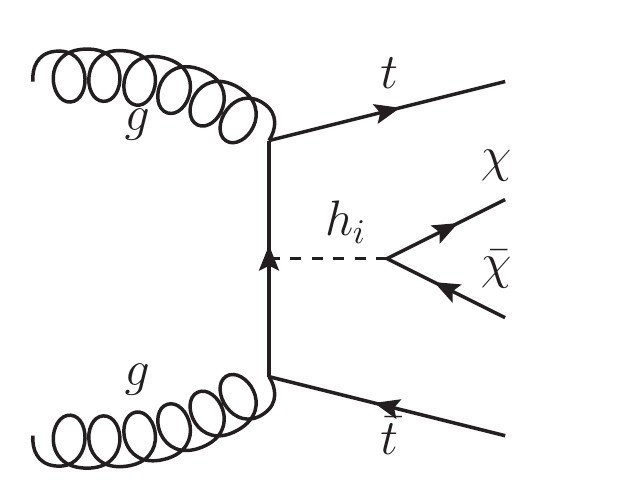}
\includegraphics[width=0.3\textwidth]{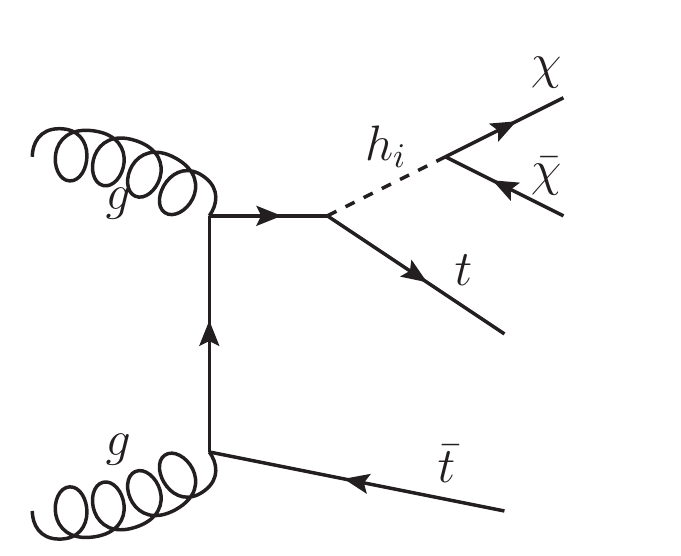}
\includegraphics[width=0.3\textwidth]{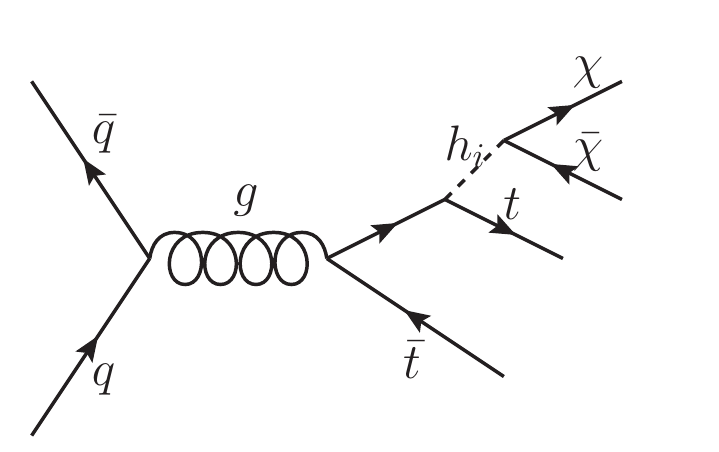}
\caption{Feynman diagrams of $pp\to t\bar{t}\chi\bar{\chi}$ in the SFDM model with mediators $h_i=h_1,h_2$.}
\label{fig:feynman_diagrams}
\end{figure}

There are two mediators in $pp\to t\bar{t}\chi\bar{\chi}$ in the SFDM model, which are shown in Fig.~\ref{fig:feynman_diagrams}. The total amplitude is then proportional to
\begin{align}
\mathcal{A}\propto g_{\chi}\sin(2\theta)(\frac{1}{\hat{s}-m_{h_1}^2+im_{h_1}\Gamma_{h_1}}
-\frac{1}{\hat{s}-m_{h_2}^2+im_{h_2}\Gamma_{h_2}}),
\label{eq:amplitude}
\end{align}
where $\hat{s}\equiv m_{\chi\bar{\chi}}^2$ with $ m_{\chi\chi}$ being the invariant mass of the DM pair
$\chi\bar{\chi}$. Therefore the diagrams with $h_1$ and the ones with $h_2$ interfere destructively for $\sqrt{\hat{s}}
>m_{h_{1}},m_{h_{2}}$ or $\sqrt{\hat{s}}<m_{h_{1}},m_{h_{2}}$ and constructively for $m_{h_{1}}
<\sqrt{\hat{s}}<m_{h_2}$ or $m_{h_{2}}<\sqrt{\hat{s}}<m_{h_1}$.
Note that in the simplfied model (see Eq.~\eqref{eq:simplified}) only  $h_2$ in Fig.~\ref{fig:feynman_diagrams} is included,
in contrast with the renormalizable and gauge-invariant SFDM model.

Depending on the relations of $m_{h_1}$, $m_{h_2}$ and $m_{\chi}$, the process $pp\to t\bar{t}
\chi\bar{\chi}$ in the SFDM model can be categorized into four cases, namely,
\begin{itemize}
\item Case A: $m_{h_1}, m_{h_2} > 2m_{\chi}$,
\item Case B: $m_{h_1} > 2m_{\chi}$ and $m_{h_2} < 2m_{\chi}$,
\item Case C: $m_{h_1} < 2m_{\chi}$ and $m_{h_2} > 2m_{\chi}$, and
\item Case D: $m_{h_1}, m_{h_2} < 2m_{\chi}$.
\end{itemize}
In the following, we will denote the cross section of diagrams with each scalar
mediator ($h_1/ h_2$) as $\sigma_{h_1}$ and $\sigma_{h_2}$, while the total cross section that
includes the interference effect between diagrams with different mediators is denoted as
$\sigma_{h_1+h_2}$.

For Case A, both $h_1$ and $h_2$ can be on-shell in DM production, so that the cross sections can
be described as
\begin{align}
\sigma_{h_1}&=c_{\theta}^2\sigma^{\text{prod}}(m_{h_1})\text{Br}(h_1\to \chi\bar{\chi}),\\
\sigma_{h_2}&=s_{\theta}^2\sigma^{\text{prod}}(m_{h_2})\text{Br}(h_2\to \chi\bar{\chi})
\end{align}
with the narrow width approximation (NWA), where $\sigma^{\text{prod}}(m_{h_i})$ denotes the
cross section of $pp\to t\bar{t}h_i$ for on-shell $h_i$, $i=1,2$ with SM couplings.
In this case, the interference effect is small unless $m_{h_2}\simeq m_{h_1}$; thus, the cross section
with two mediators is approximately equal to the sum of the cross sections with one mediator:
\begin{align}
\sigma_{h_1+h_2}\simeq \sigma_{h_1}+\sigma_{h_2}.
\end{align}

\begin{figure}[!thb]
\includegraphics[width=0.40\textwidth]
{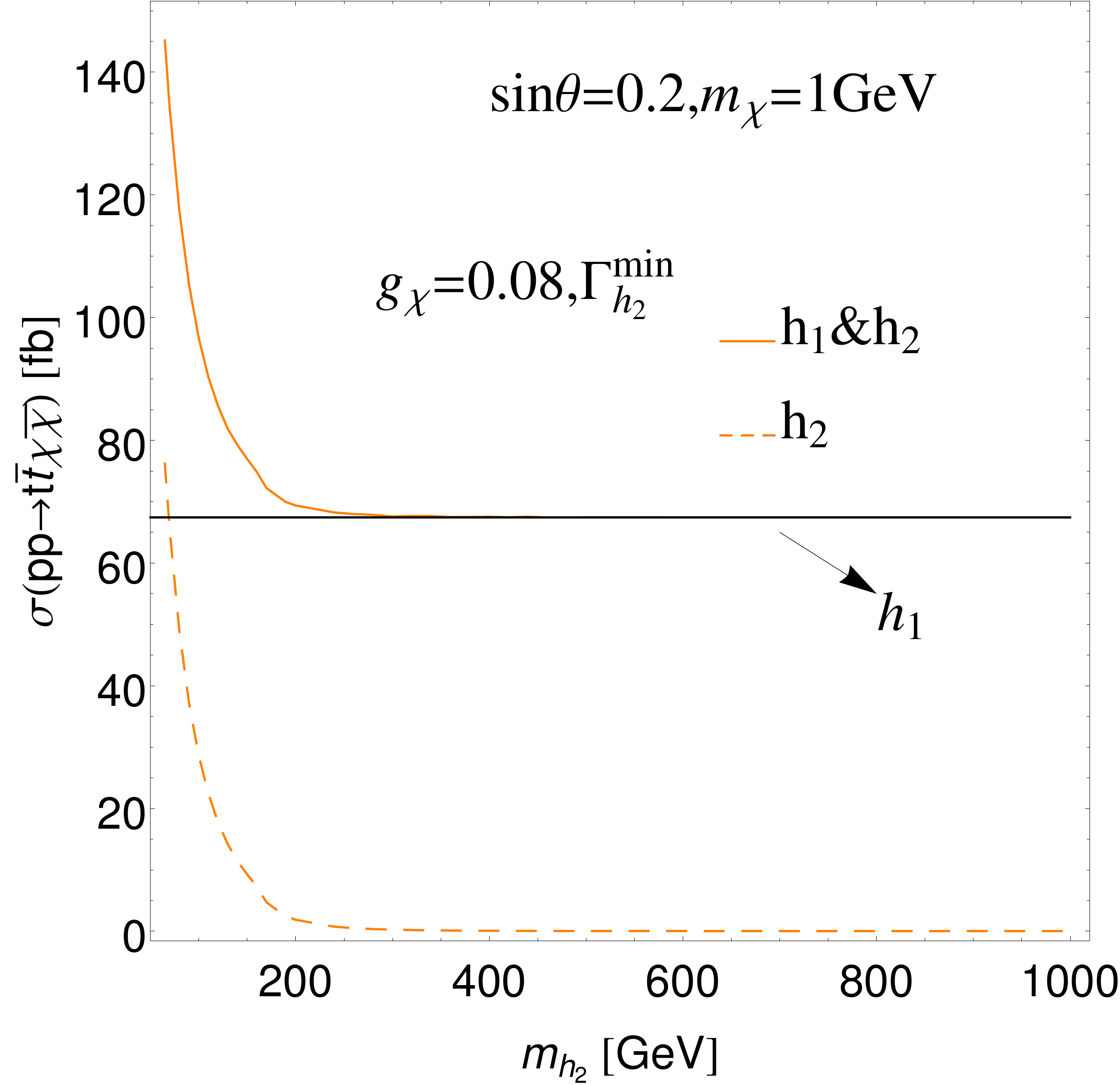}
\includegraphics[width=0.40\textwidth]
{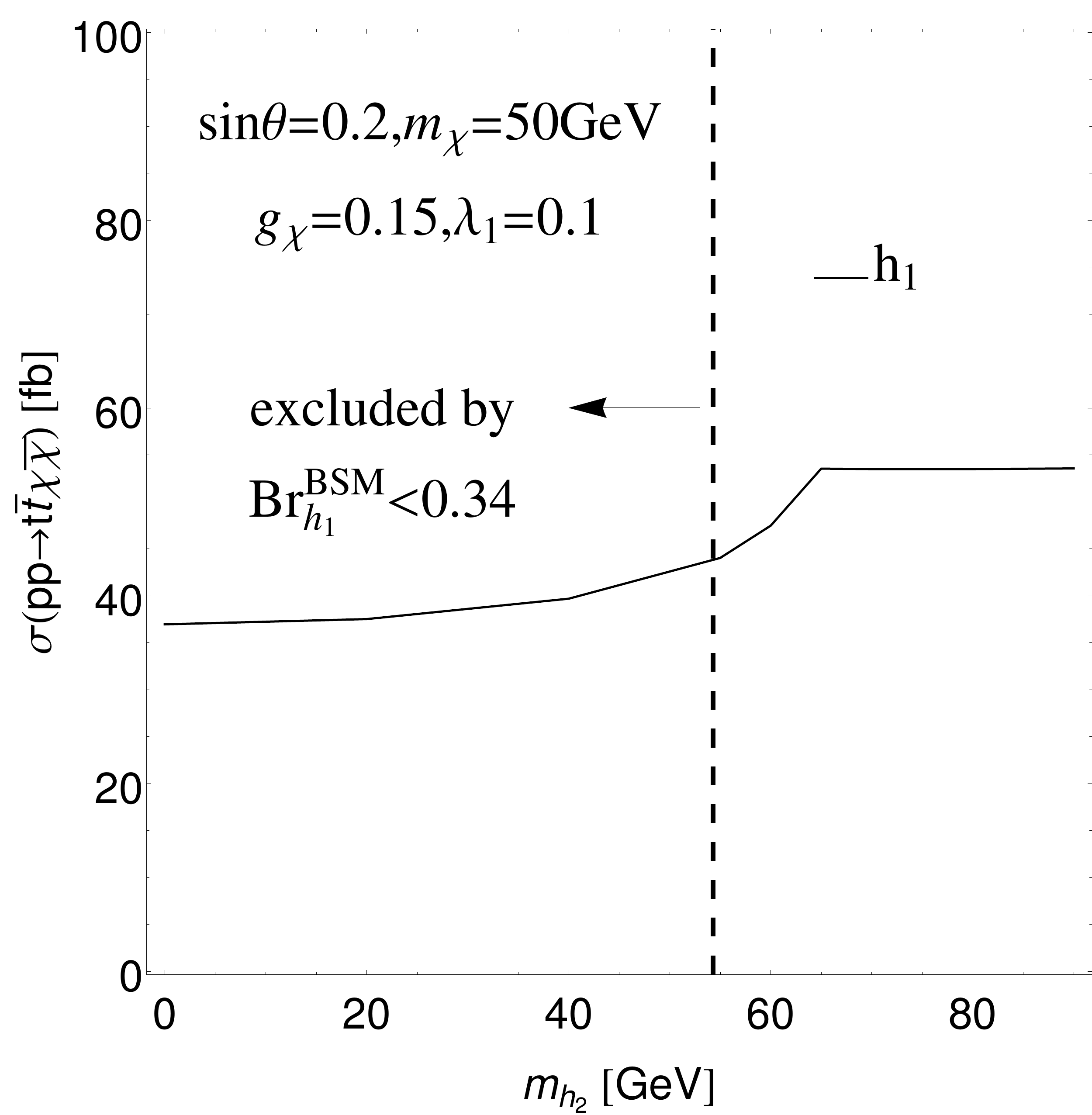}
\caption{Cross sections of $pp\to t\bar{t}\chi\bar{\chi}$ at the 13 TeV LHC for Case A (left panel)
with $g_{\chi}=0.08$, $m_{\chi}=1~\text{GeV}$ and $\Gamma_{h_2}=\h2min$ and for Case B
(right panel) with $g_{\chi}=0.15$, $\lambda_{1}=0.1$ and $m_{\chi}=50~\text{GeV}$, where
$m_{h_2}\lesssim 54.3\gev$ is excluded by the BSM decay branching ratio of $h_1$.}
\label{fig:LOxsec_AB}
\end{figure}

In the left panel of Fig.~\ref{fig:LOxsec_AB}, we show the leading-order (LO) cross section of
$pp\to t\bar{t}\chi\bar{\chi}$ at the 13 TeV LHC for Case A with $g_{\chi}=0.08$, $m_{\chi}=
1~\text{GeV}$, $\Gamma_{h_2}=\h2min$ and $m_{h_2}\gtrsim 65\gev$ satisfying the constraint
from the invisible decay branching ratio of $h_1$.  We  find that the $h_2$ provides a larger cross
section than the $h_1$ when $m_{h_2}\lesssim 70\gev$. With increasing $m_{h_2}$, the contribution
from $h_2$ decreases dramatically and becomes negligible for $m_{h_2}\gtrsim 300\gev$, in which
scenario the process $pp\to t\bar{t}\chi\bar{\chi}$ is effectively described by a single mediator $h_1$.
Note that here we assume $\Gamma_{h_2}=\h2min$. If we consider the decay of $h_2$ into
$h_1h_1$, which is possible if $m_{h_2}>2m_{h_1}$, the branching ratio of $h_2\to \chi\bar{\chi}$
will be suppressed. On the other hand, $h_1$ can also decay into $h_2h_2$ if $m_{h_2}<m_{h_1}/2$
so that the branching ratio of $h_1\to \chi\bar{\chi}$ is suppressed. This fact will bring the $m_{h_2}$
dependence to the cross section of $\sigma_{h_1}$, which is not shown in the
Fig.~\ref{fig:LOxsec_AB}.

For Case B, only $h_1$ can be on-shell so that the contribution of $h_1$ is dominant and
\begin{align}
\sigma_{h_1+h_2}\simeq \sigma_{h_1}.
\end{align}
In the right panel of Fig.~\ref{fig:LOxsec_AB}, we show the cross section of
$pp\to t\bar{t}\chi\bar{\chi}$ at the 13 TeV LHC for Case B with $g_{\chi}=0.15$, $\lambda_1=0.1$, and
$m_{\chi}=50~\text{GeV}$. Similar to Case A, the cross section $\sigma_{h_1}$ depends on
$m_{h_2}$ in the region of $m_{h_2}<m_{h_1}/2$. Considering the constraint from the BSM decay
branching ratio of $h_1$, $\text{Br}^{\text{BSM}}_{h_1}<0.34$~\cite{Khachatryan:2016vau},
$h_2$ with $m_{h_2}\lesssim 54.3\gev$ is excluded for our parameter choice.

\begin{figure}[!thb]
\includegraphics[width=0.40\textwidth]
{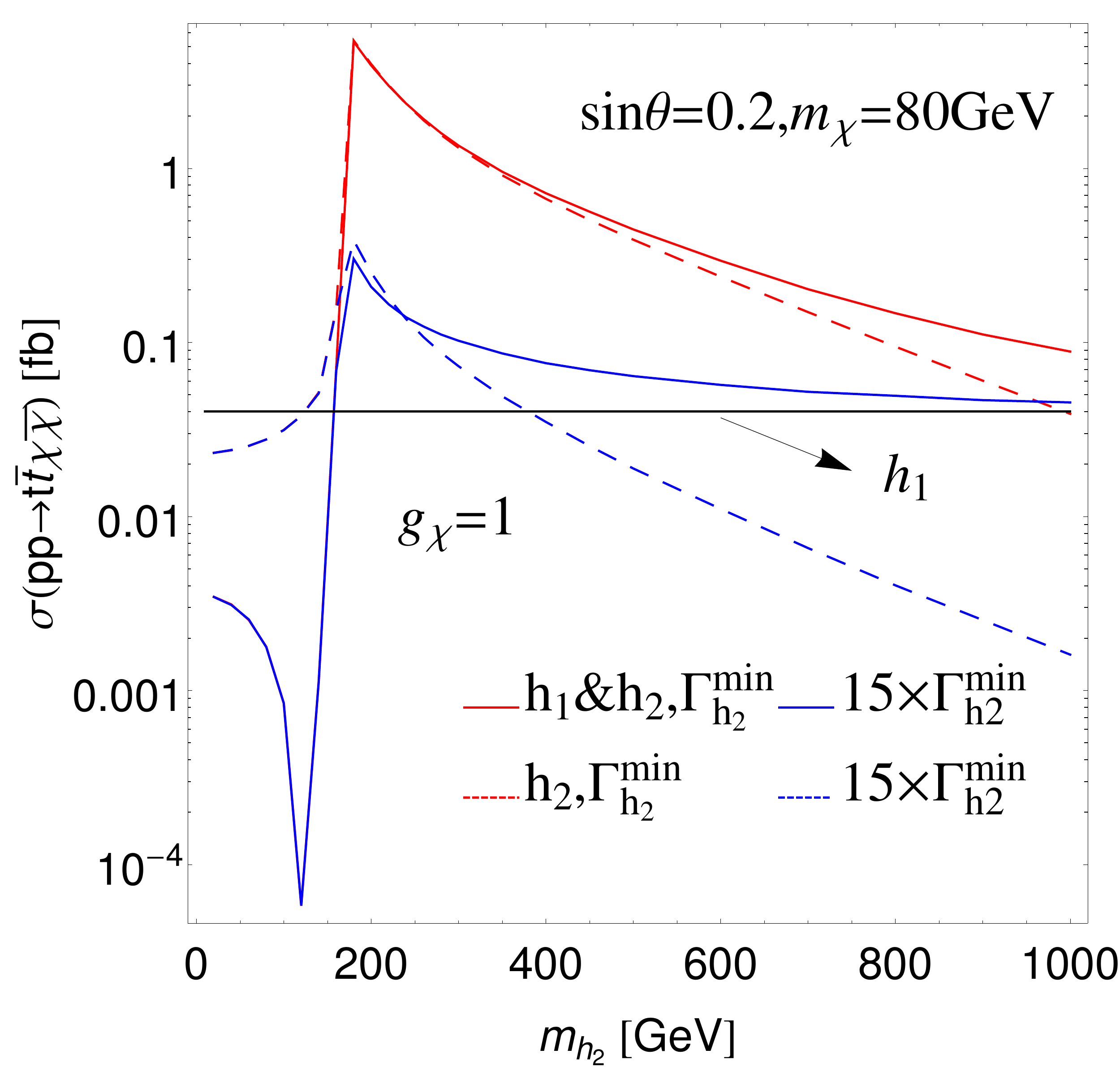}
\includegraphics[width=0.40\textwidth]
{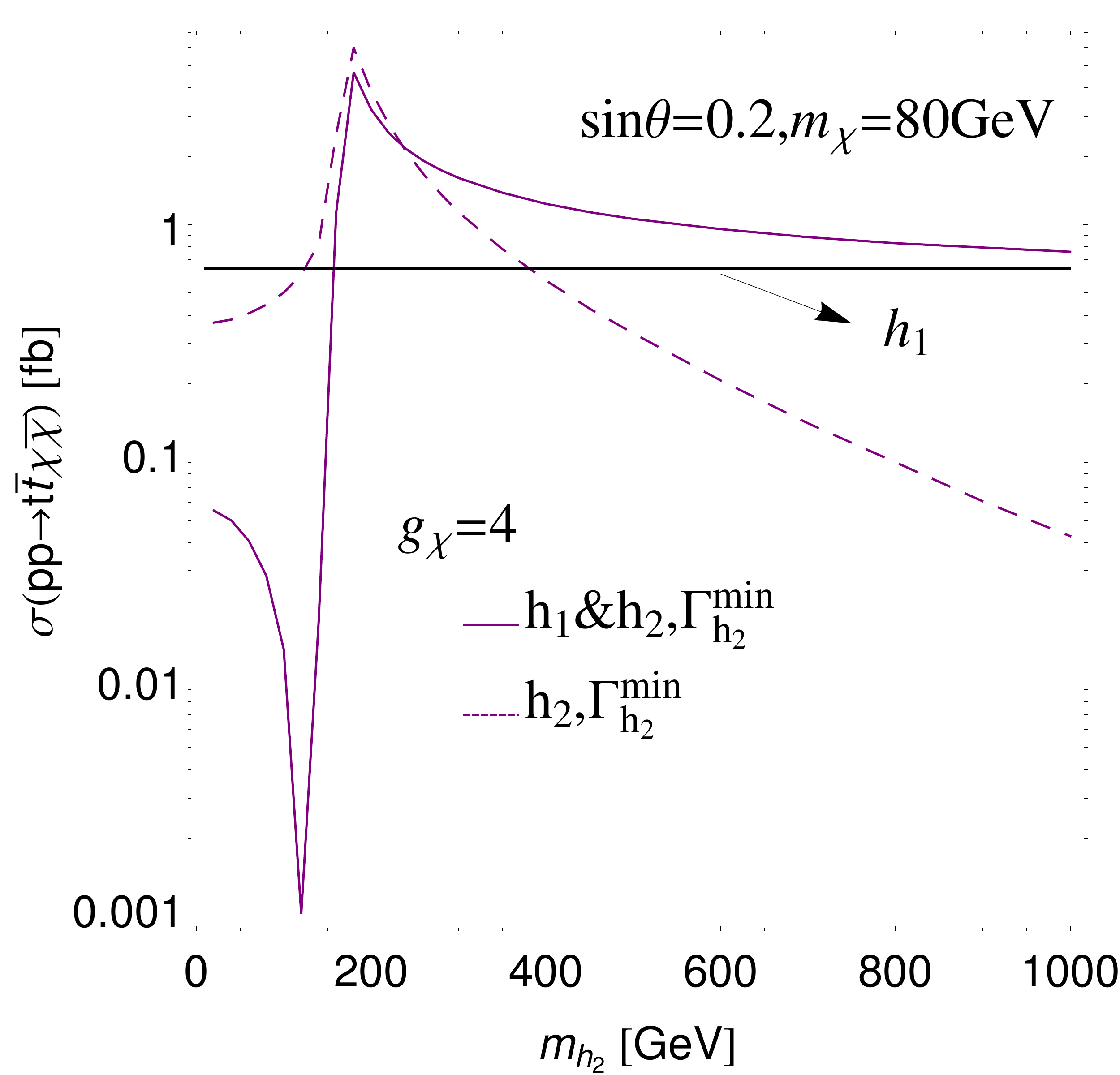}
\caption{Cross sections of $pp\to t\bar{t}\chi\bar{\chi}$ at the 13 TeV LHC for Case C and Case D.
Left panel: $g_{\chi}=1$, $m_{\chi}=80~\text{GeV}$ and $\Gamma_{h_2}=\h2min,15\times\h2min$;
right panel: $g_{\chi}=4$, $m_{\chi}=80~\text{GeV}$ and $\Gamma_{h_2}=\h2min$.}
\label{fig:LOxsec_CD}
\end{figure}

For Case C, only $h_2$ can be on-shell. If $\Gamma_{h_2}/m_{h_2}\ll 1$, which implies that the
NWA can be applied, one obtains
\begin{align}
\label{caseC:eq:h1}
\sigma_{h_1}&\propto s_{\theta}^2c_{\theta}^2g_{\chi}^2,\\
\label{caseC:eq:h2}
\sigma_{h_2}&=s_{\theta}^2\sigma^{\text{prod}}(m_{h_2})\text{Br}(h_2\to \chi\bar{\chi}).
\end{align}
If $\Gamma_{h_2}=\h2min$, as shown in Fig.~\ref{fig:branching_ratios} the decay branching ratio
$\text{Br}(h_2\to \chi\bar{\chi})$ is relatively large for $g_{\chi}\sim 1$, which leads to
$\sigma_{h_2}\gg\sigma_{h_1}$, so that the simplified model with single mediator $h_2$ can
describes the SFDM model approximately. However, there are key differences between this case
and Case B.
Firstly, confronted with the measurements of the Higgs invisible decay branching ratio the coupling
$g_{\chi}$ is severely constrained for Case B, while there is no such constraint on $g_{\chi}$ for
Case C due to $m_{h_1}<2m_{\chi}$. Thus the cross section with off-shell $h_1$ can be significantly
enhanced for a large  $g_{\chi}\sim 5$~\cite{Haisch:2015ioa,Arina:2016cqj} as in Eq.~\eqref{caseC:eq:h1}.
Secondly, unlike $h_1$ there is no direct experimental constraint on $\Gamma_{h_2}$. So if there exist other decay channels of $h_2$, for example $h_2$ decays into extra dark sector particles~\cite{Baek:2014kna}, the total width of $h_2$ can be significantly enhanced as compared to $\h2min$. Consequently, the branching
ratio of $h_2$ into $\chi\bar{\chi}$ is possibly small. Thus although $h_1$ is off-shell in DM production
for Case C, its contribution should be taken into account, which however has not been paid much
attention to. It should be noted that the wide width of $h_2$ may make the NWA in
Eq.~\eqref{caseC:eq:h2} invalid.

For Case D, both $h_1$ and $h_2$ are off-shell, so that the cross sections
\begin{align}
\sigma_{h_1}&\propto s_{\theta}^2c_{\theta}^2 g_{\chi}^2, \\
\sigma_{h_2}&\propto s_{\theta}^2c_{\theta}^2 g_{\chi}^2
\end{align}
are small but the interference effect between diagrams with different mediators in Fig.~\ref{fig:feynman_diagrams}
is significant and always destructive, as can be seen from Eq.~\eqref{eq:amplitude}.

\begin{figure}[!thb]
\includegraphics[width=0.3\textwidth]{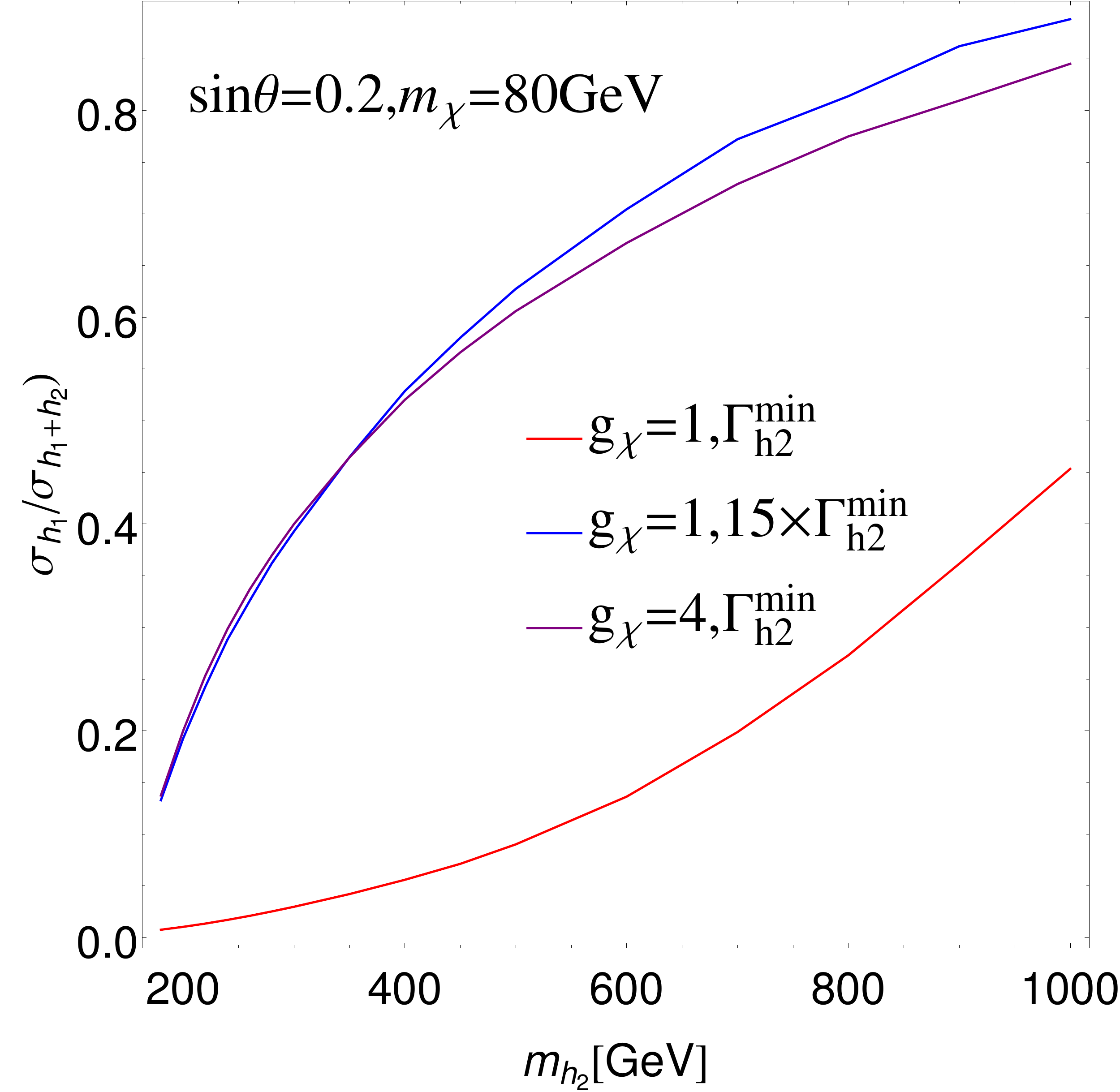}
\includegraphics[width=0.3\textwidth]{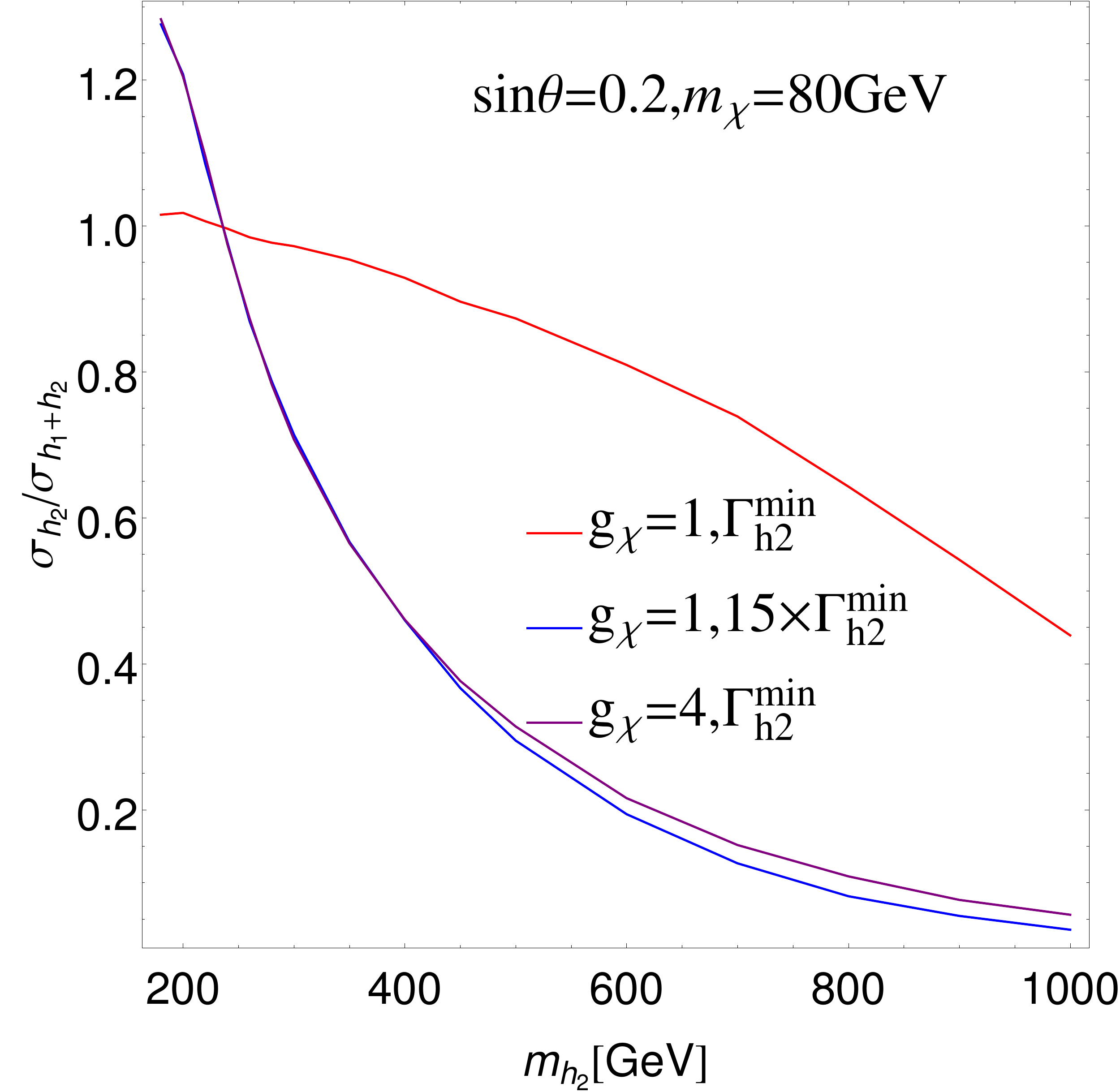}
\includegraphics[width=0.3\textwidth]{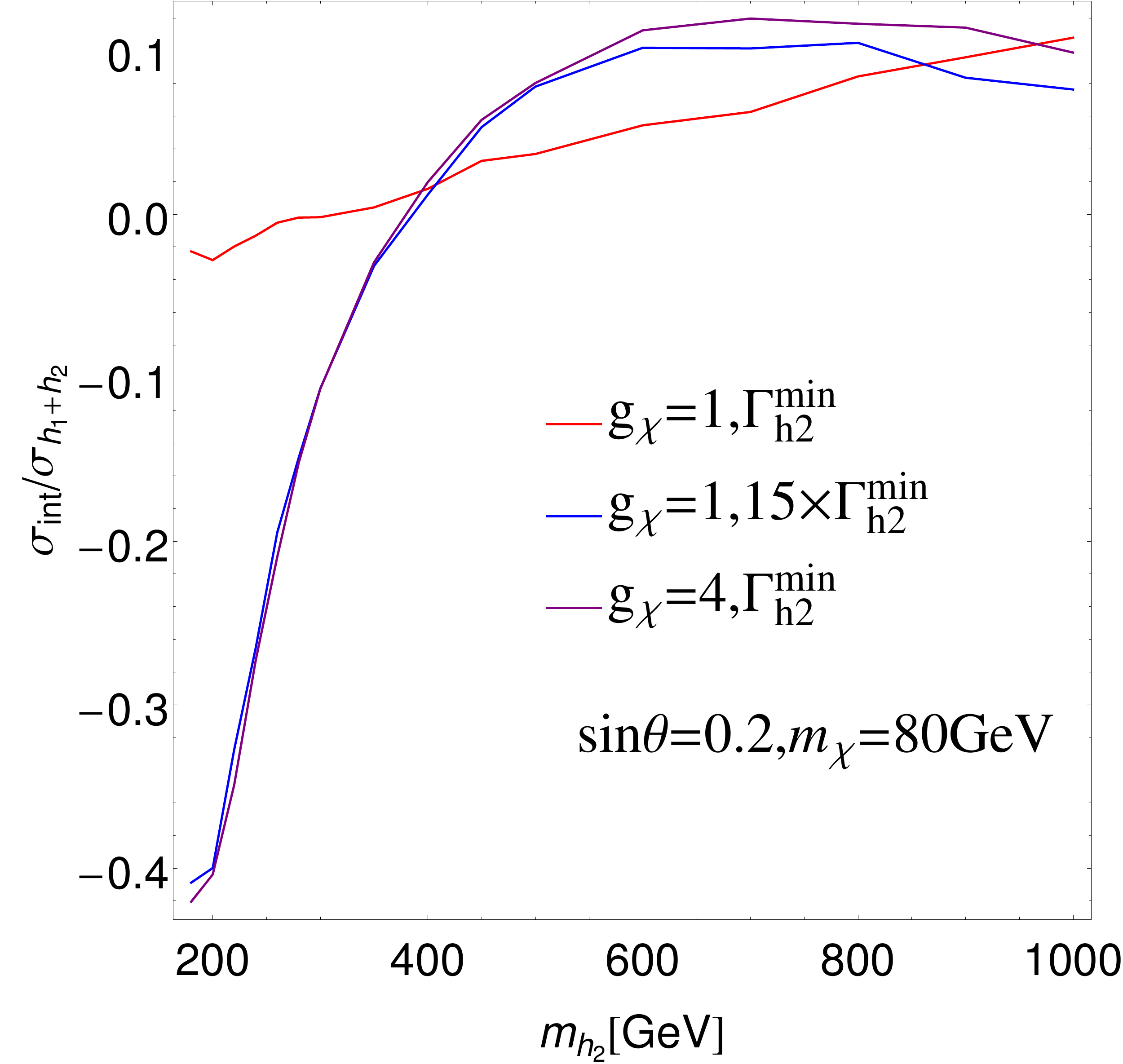}
\caption{The relative contributions of mediators $h_1$, $h_2$ and their interference to the total
cross section for Case C.}
\label{fig:ratio}
\end{figure}

To consider the wide width effects, we assume that the total width of $h_2$ is rescaled by a factor
of 15 irrespective of its mass, i.e., $\Gamma_{h_2}=15\times\h2min$, which satisfies
$\Gamma_{h_2}/m_{h_2}<1$ as shown in Fig.~\ref{fig:width_h2}. Then, the cross section of
$pp\to t\bar{t}\chi\bar{\chi}$ with $h_2$ mediation is reduced by a factor of about $15$. On the other
hand, a larger coupling $g_{\chi}$ can also increase the $h_2$ decay width as well as enhance the
$\sigma_{h_1}$. To compare with the simplified model study in Refs.~\cite{Haisch:2015ioa,
Arina:2016cqj}, we choose $g_{\chi}=4$. In Fig.~\ref{fig:LOxsec_CD}, we show the cross sections
of $pp\to t\bar{t}\chi\bar{\chi}$ at the 13 TeV LHC for Case C and Case D with $g_{\chi}=1$,
$m_{\chi}=80~\text{GeV}$, and $\Gamma_{h_2}=\h2min(15\times\h2min)$ in the left panel and
$g_{\chi}=4$, $m_{\chi}=80~\text{GeV}$, and $\Gamma_{h_2}=\h2min$ in the right panel.

We find that $\sigma_{h_1+h_2}<\sigma_{h_1}+\sigma_{h_2}$ in the region of $m_{h_2}\lesssim
2m_{\chi}$ (Case D) irrespective of $g_{\chi}$ and $\Gamma_{h_2}$. This is due to the destructive
interference between diagrams with $h_1$ and $h_2$ in case D. The destructive interference effect
is most significant at $m_{h_2}\simeq m_{h_1}$.

For $m_{h_2}\gtrsim 2m_{\chi}$ (Case C), the contributions of $h_1$ and $h_2$ depend on the
coupling $g_{\chi}$ and the total width of $h_2$ as we discussed above. For illustration, in
Fig.~\ref{fig:ratio} we show the relative contributions of mediator $h_1$, mediator $h_2$ and their
interference contributions to the total cross sections denoted as $\sigma_{h_1}/\sigma_{h_1+h_2}$,
$\sigma_{h_2}/\sigma_{h_1+h_2}$ and $\sigma_{\text{int}}/\sigma_{h_1+h_2}$. For $g_{\chi}=1$
and $\Gamma_{h_2}=\h2min$, it is observed that $\sigma_{h_2}/\sigma_{h_1+h_2}$ is approximately
equal to 1 for $m_{h_2}\simeq 200\gev$ and decreases with larger $m_{h_2}$. On the other hand,
both $\sigma_{h_1}/\sigma_{h_1+h_2}$ and $\sigma_{\text{int}}/\sigma_{h_1+h_2}$ become more
important with the increase of $m_{h_2}$. For example, $\sigma_{h_1}:\sigma_{h_2}:
\sigma_{\text{int}}=0.087:0.87:0.037$ for $m_{h_2}=500\gev$ and $\sigma_{h_1}:\sigma_{h_2}:
\sigma_{\text{int}}=0.45:0.45:0.1$ for $m_{h_2}=1000\gev$. The interference effect is destructive
in the region of $2m_{\chi}\lesssim m_{h_2}\lesssim 380\gev$ and constructive for $m_{h_2}\gtrsim
380\gev$. The impact of $h_1$ becomes even more significant for $\Gamma_{h_2}=15\times\h2min$
or $g_{\chi}=4$.

\begin{figure}[!thb]
\includegraphics[width=0.32\textwidth]{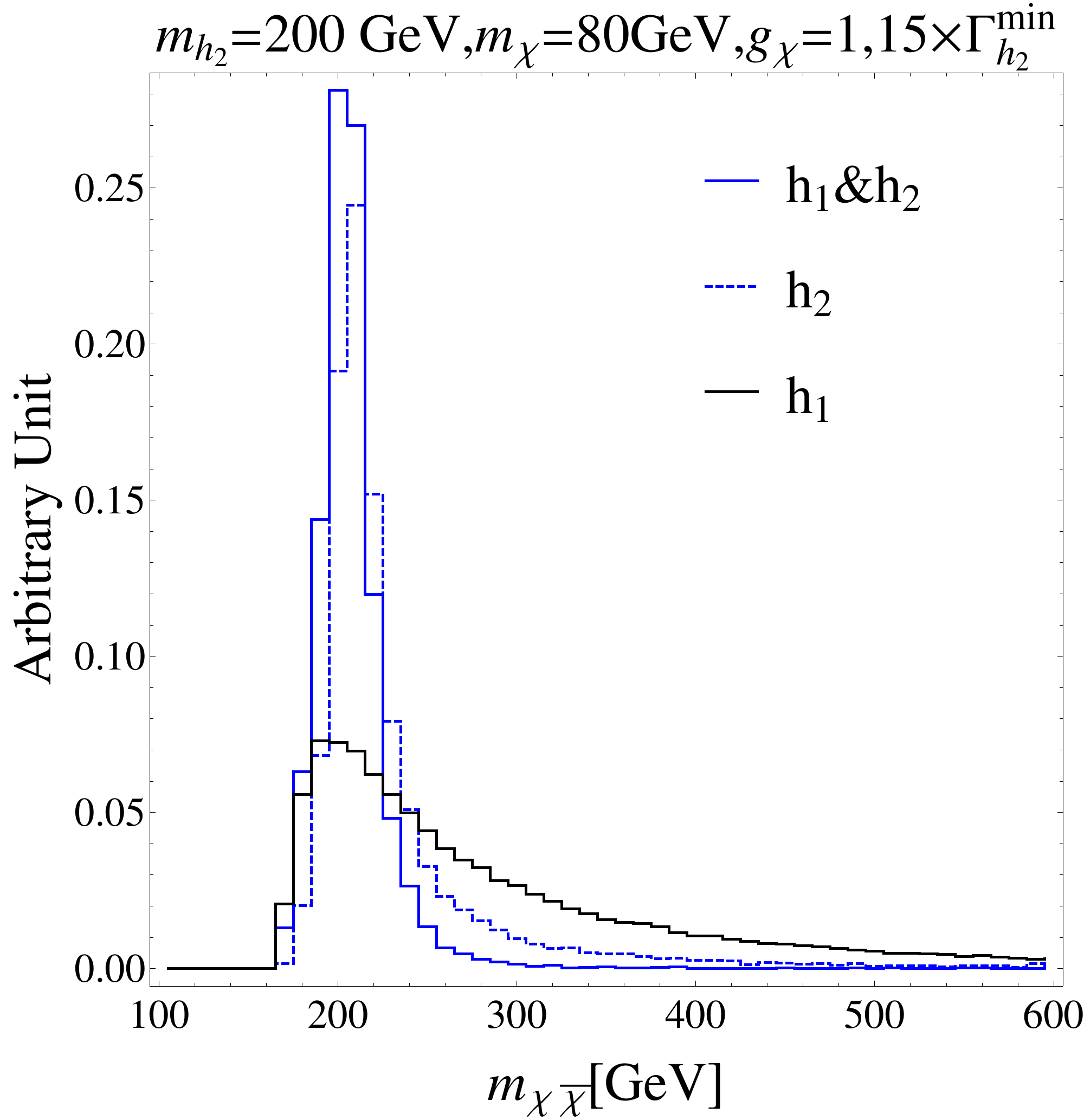}
\includegraphics[width=0.32\textwidth]{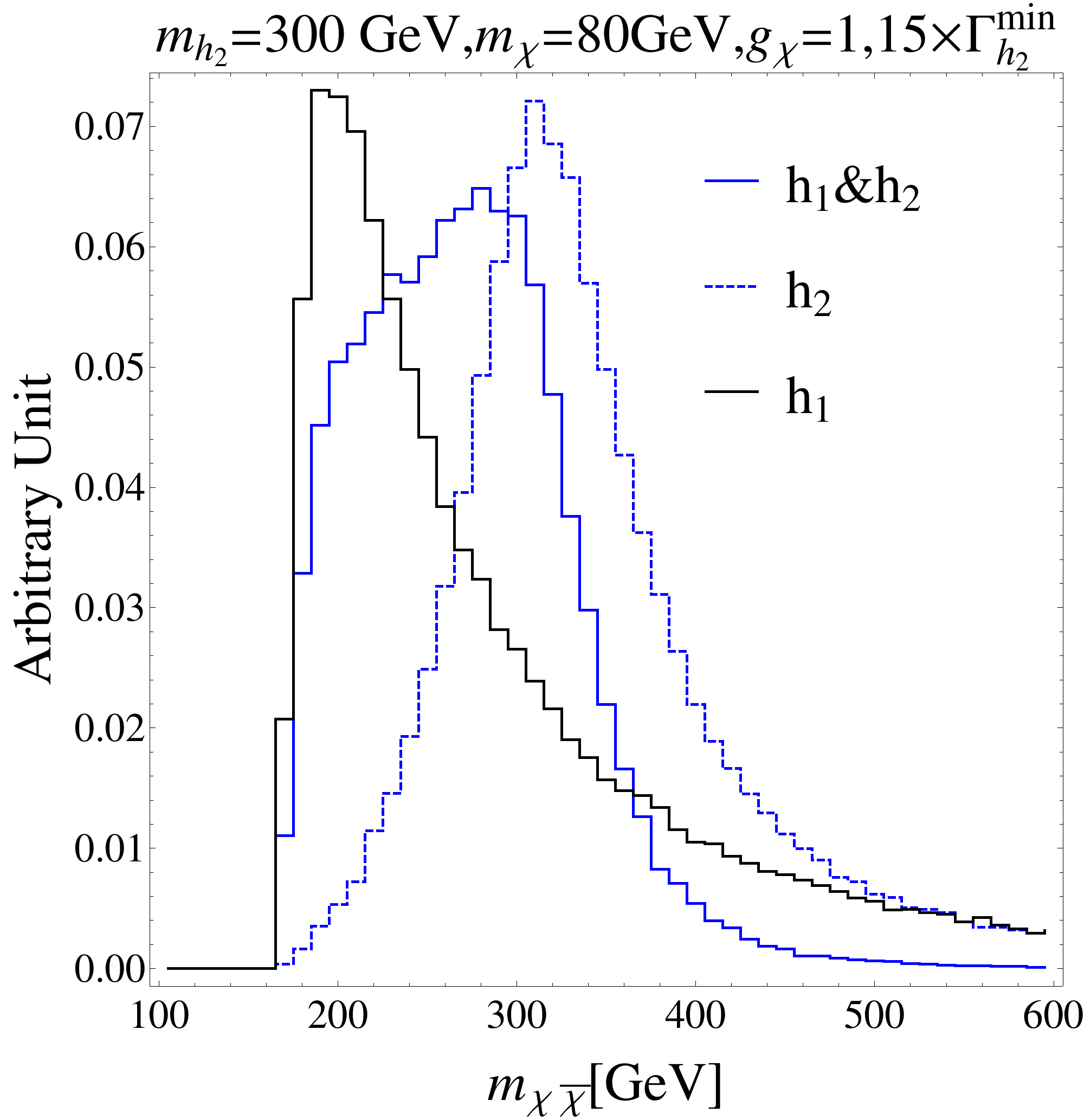}
\includegraphics[width=0.32\textwidth]{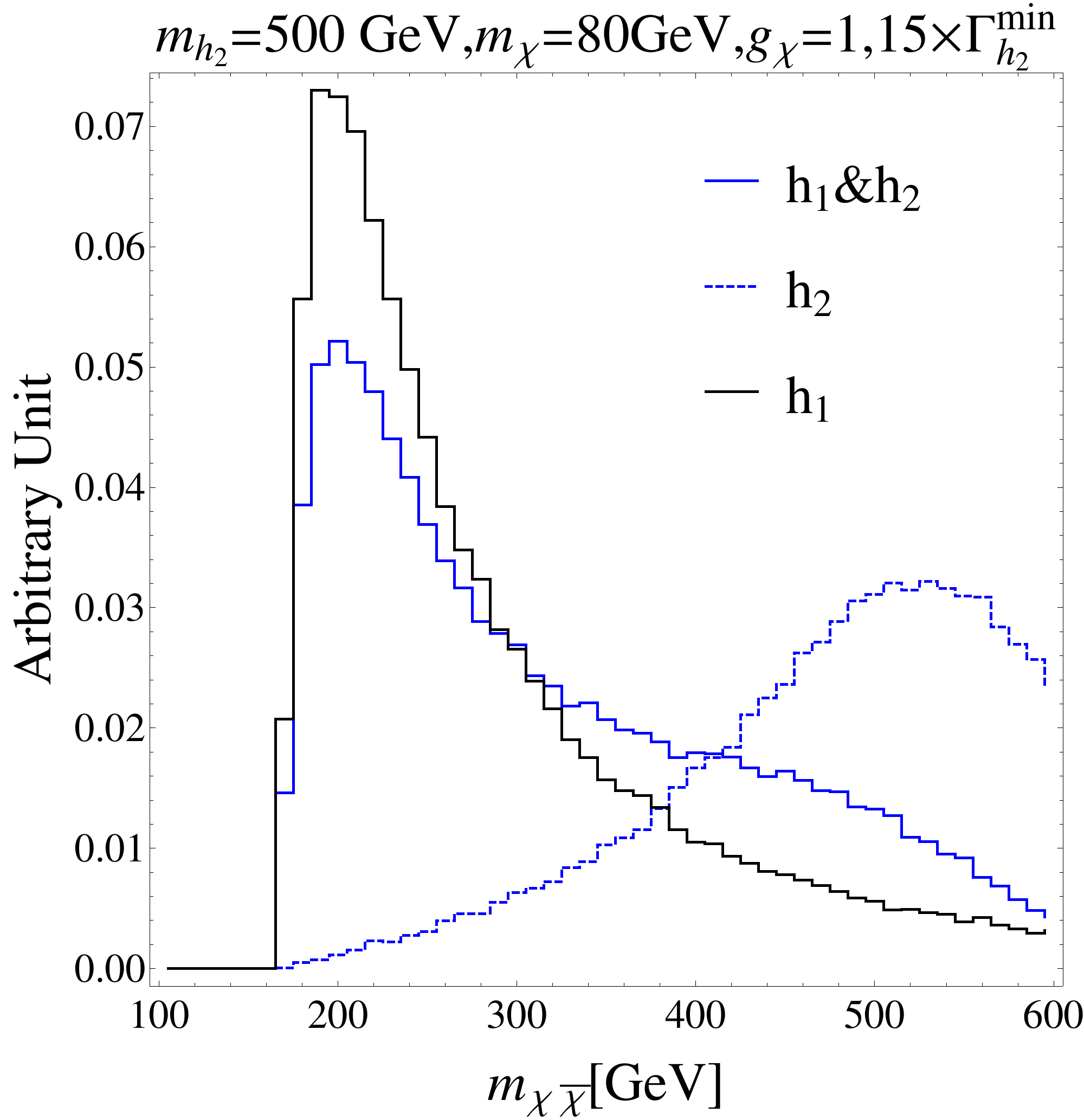}
\caption{The distributions of $m_{\chi\bar{\chi}}$ with mediators $h_1$ and $h_2$ for Case C with
$m_{\chi}=80\gev$, $m_{h_2}=200,300,500\gev$, $g_{\chi}=1$ and
$\Gamma_{h_2}=15\times\h2min$.}
\label{fig:mn1n1}
\end{figure}

Until now we have only discussed the impact of the mediator $h_1$ on the total cross section, in
which the $m_{\chi\bar{\chi}}$ dependence has been integrated out. From Eq.~\eqref{eq:amplitude},
we see that the interference effect depends on the interplay of $m_{\chi\bar{\chi}}^2-m_{h_1}^2$ with
$m_{\chi\bar{\chi}}^2-m_{h_2}^2$. For example, the interference is constructive (destructive) for
$m_{\chi\bar{\chi}}^2-m_{h_2}^2<(>)0$ for Case C. To study the impact of $h_1$ on the differential
cross section, the invariant mass distributions of $\chi\bar{\chi}$ are displayed with $m_{\chi}=80\gev$,
$m_{h_2}=200$, $300$, $500\gev$, $g_{\chi}=1$ and $\Gamma_{h_2}=15\times\h2min$ (similar for
$g_{\chi}=4$ and $\Gamma_{h_2}=\h2min$) in Fig.~\ref{fig:mn1n1}. Unlike the distribution with
on-shell $h_2$ (blue dashed curve), which centers around $m_{\chi\bar{\chi}}\sim m_{h_2}$, the
distribution with off-shell $h_1$ (black curve) decreases rapidly with increasing $m_{\chi\bar{\chi}}$.
As a result, the distribution with two mediators tends to be softer than that in the simplified model with
$h_2$. Moreover, the distribution with two mediators is asymmetric around $m_{h_2}$: More events
fall into region of $m_{\chi\bar{\chi}}^2-m_{h_1}^2>0$ for $m_{h_2}=200$ and $300\gev$, while fewer
are in the region of $m_{\chi\bar{\chi}}^2-m_{h_1}^2>0$ for $m_{h_2}=500\gev$. After integrating over
$m_{\chi\bar{\chi}}$, the interference effect in the total cross section is destructive for $m_{h_2}
=200$ and $300\gev$ and constructive for $m_{h_2}=500\gev$ as shown in the right panel of
Fig.~\ref{fig:ratio}.

\begin{figure}[!thb]
\includegraphics[width=0.3\textwidth]{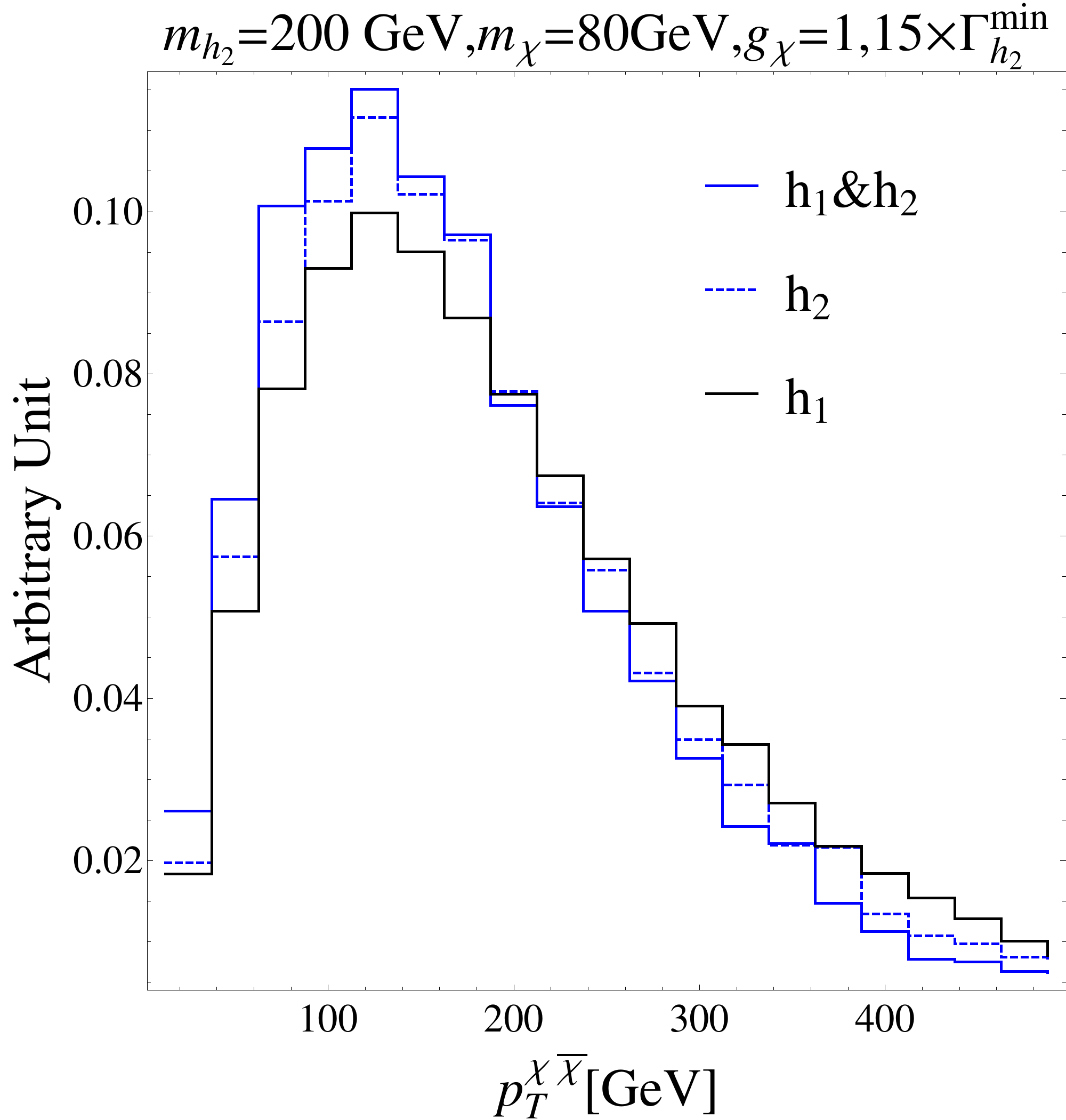}
\includegraphics[width=0.3\textwidth]{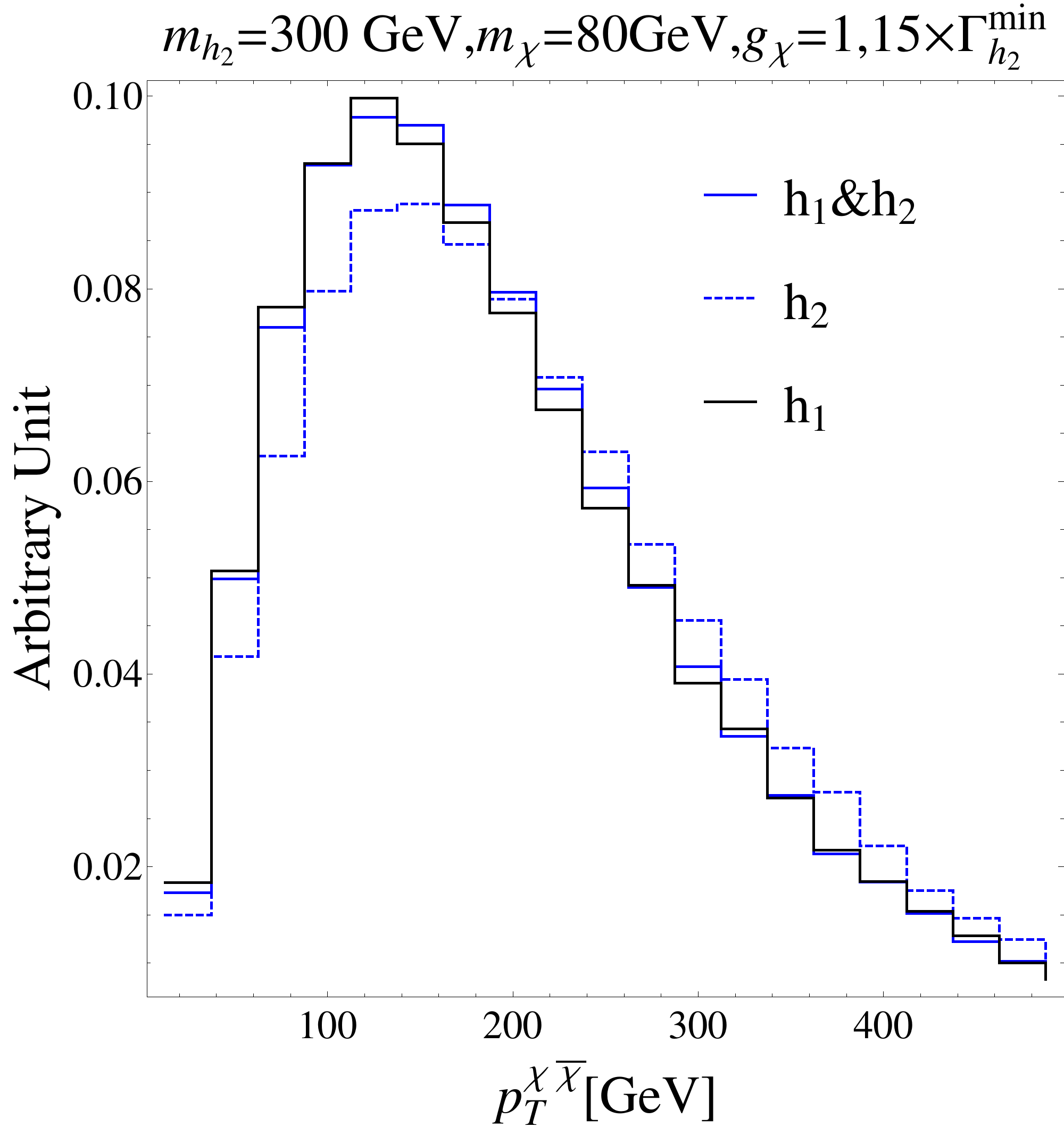}
\includegraphics[width=0.3\textwidth]{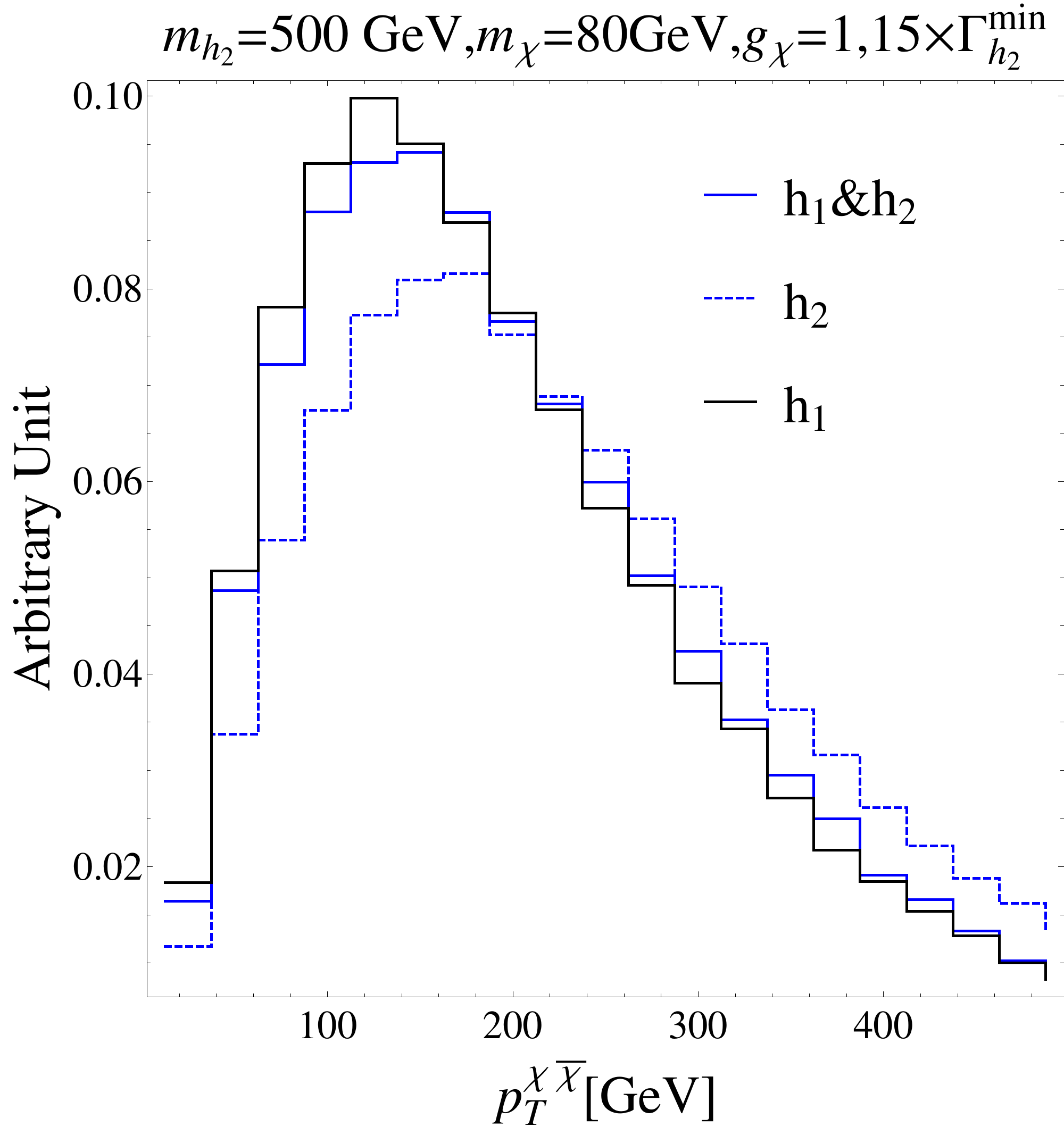}
\caption{The parton-level distributions of $p_{T}^{\chi\bar{\chi}}$ for Case C with $m_{\chi}=80\gev$,
$m_{h_2}=200$, $300$, $500\gev$, $g_{\chi}=1$ and $\Gamma_{h_2}=15\times\h2min$.}
\label{fig:ptn1n1}
\end{figure}

Experimentally, the invariant mass of $\chi\bar{\chi}$ cannot be reconstructed directly. However, its
feature can be reflected in the distribution of the transverse momentum of $\chi\bar{\chi}$ (missing
transverse momentum), i.e., $p_{T}^{\chi\bar{\chi}}$. In Fig.~\ref{fig:ptn1n1} we show the
parton-level distributions of $p_{T}^{\chi\bar{\chi}}$ for Case C with $m_{\chi}=80\gev$, $m_{h_2}
=200,300,500\gev$, $g_{\chi}=1$ and $\Gamma_{h_2}=15\times\h2min$ (similar for $g_{\chi}=4$
and $\Gamma_{h_2}=\h2min$). One can obtain that $p_{T}^{\chi\bar{\chi}}$ distribution in the SFDM
model with two mediators is always softer than that in the simplified model with $h_2$, the effect of
which on the cut efficiency of a concrete experimental search will be discussed in
Section~\ref{sec:limits}.

\section{Impact on the upper limits}
\label{sec:limits}

Having observed the distinct difference between the total cross sections and differential distributions
of $pp\to t\bar{t}\chi\bar{\chi}$ with two mediators and with one mediator, we will investigate the impact of
the Higgs boson $h_1$ on the 95\% C.L. upper limits of searches for DM produced with top quark
pair at the 13~TeV LHC, which have been performed by the CMS and ATLAS Collaborations in the hadronic,
semileptonic and dileptonic channels~\cite{CMS:2016mxc,Sirunyan:2017xgm,CMS:2018jww,Aaboud:2017rzf}. Similar to Ref.~\cite{Pinna:2017tay}, we closely follow the CMS analyses~\cite{CMS:2016mxc,Sirunyan:2017xgm,CMS:2018jww} and concentrate on the hadronic and
semileptonic channels since the dileptonic are typically less sensitive in these analyses. On the other hand, although recent search using events with $35.9\fbi$~\cite{CMS:2018jww} has improved the upper limit significantly as compared to that with $2.2\fbi$~\cite{CMS:2016mxc,Sirunyan:2017xgm}, the background event numbers in signal regions are not provided~\cite{CMS:2018jww}. Therefore we take the strategy that we first recast the results with $2.2\fbi$ and then project them to the integrated luminosity
of $36\fbi$ with the assumption that the signal and background uncertainties scale as the integrated
luminosity and the square root of the integrated luminosity, respectively. Another reason for projection is that a multivariate discriminant ``resolved top tagger'' was used in the analysis~\cite{CMS:2018jww} without giving any details for recasting.~\footnote{Due to these limitations, it is not possible to compare our upper limits with $36\fbi$ to Ref.~\cite{CMS:2018jww} directly. However we have validated our result by simulating the signal process in the simplified model with $g_{\chi}=g_q=1$ and $(m_{\chi},m_{\phi})=(1\gev,100\gev)$, $(1\gev,200\gev)$ and $(50\gev,300\gev)$. The discrepancies between our results and Refs.~\cite{Pinna:2017tay,CMS:2016mxc,Sirunyan:2017xgm} are 13\% and 23\%  in the inclusive hadronic and
semileptonic channels, respectively, which indicates that our recast is reasonable.} 
In our analysis, we only consider the inclusive hadronic channel without using ``resolved top tagger'' similar to Ref.~\cite{Pinna:2017tay}.

We generate the signal process $pp\to t\bar{t}\chi\bar{\chi}$ using \texttt{MG5\_aMC@NLO v2.4.3}
~\cite{Alwall:2014hca} at leading order in the 5-flavor scheme. The parton-level events are then
passed to \texttt{Pythia6}~\cite{Sjostrand:2006za} for parton showering and hadronization.
The detector effects are included by using \texttt{Delphes3 v3.3.3}~\cite{deFavereau:2013fsa} and
\texttt{Fastjet}~\cite{Cacciari:2011ma} packages, in which jets are clustered using the anti-$k_t$ algorithm with $R=0.4$.   The $b$-tagging efficiency is 0.6, while the $c$ quark and light quark
faking rates are 0.07 and 0.01, respectively~\cite{Pinna:2017tay}.

Having imposed the selection cuts~\cite{CMS:2016mxc,Sirunyan:2017xgm}, the number of signal
events with integrated luminosity $\mathcal{L}$ is then
\begin{align}
\label{eq:signal_events}
\mathcal{N}_{\text{signal}}&=\sigma_{\text{before cut}}\times \epsilon \times \mathcal{L},
\end{align}
where $\sigma_{\text{before cut}}$ denotes the cross section of $pp\to t\bar{t}\chi\bar{\chi}$
without any cut and $\epsilon$ denotes the cut efficiency in the hadronic or semileptonic channel.

The LHC search sensitivity to our models can be measured by the signal strength $\mu=\sigma
/ \sigma_{\text{th}}$,  where $\sigma$ denotes the observed production cross section of
$pp\to t\bar{t}\chi\bar{\chi}$ at the 13~TeV LHC and $\sigma_{\text{th}}$ is the theoretical signal
cross section. The 95\% C.L. upper limit on $\mu$ is investigated using the CLs method
~\cite{Read:2002hq,Junk:1999kv,Cowan:2010js}, with inputs of production cross section and cut
efficiencies of a concrete model from our simulation as well as the number of background events
and their uncertainties provided in the experimental papers~\cite{Sirunyan:2017xgm,CMS:2016mxc}.

Then we will investigate the impact of the mediator $h_1$ on the upper limits $\mu$ in the SFDM
model. From Eq.~\eqref{eq:signal_events}, we know that the number of signal events after selections
depend on the total cross section of $pp\to t\bar{t}\chi\bar{\chi}$ as well as the cut efficiency.
As having been discussed in Section~\ref{sec:xsection}, for Case A and Case B, the signal production
is dominated by the process with mediator $h_1$. So adding $h_1$ will change both the production
rate and the final state kinematics (i.e., cut efficiency) significantly.  On the other hand, for Case C
and Case D, both $h_1$ and $h_2$ have an impact on the signal production cross section and cut
efficiency.

\begin{figure}[!thb]
\includegraphics[width=0.31\textwidth]{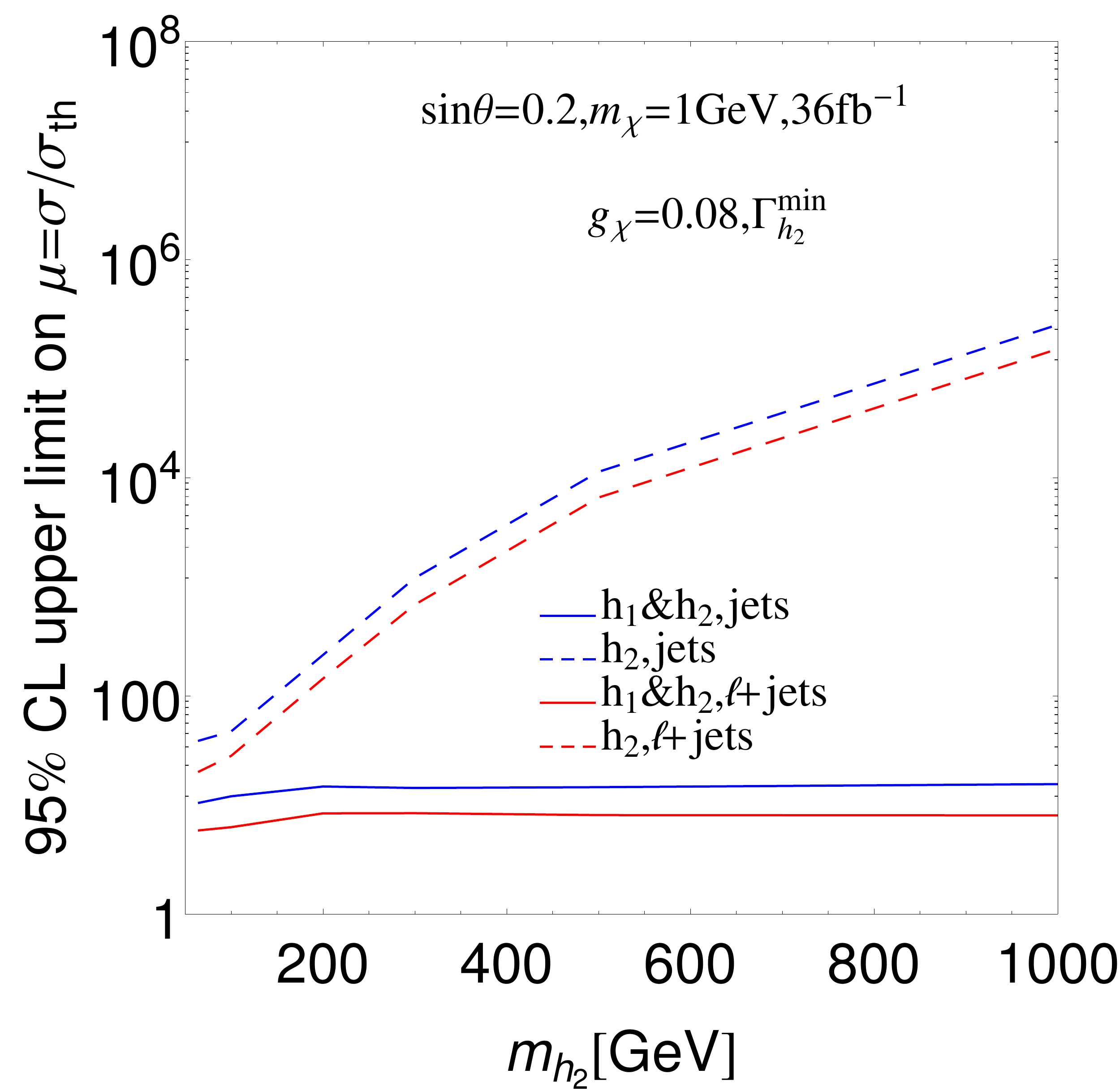}
\includegraphics[width=0.33\textwidth]{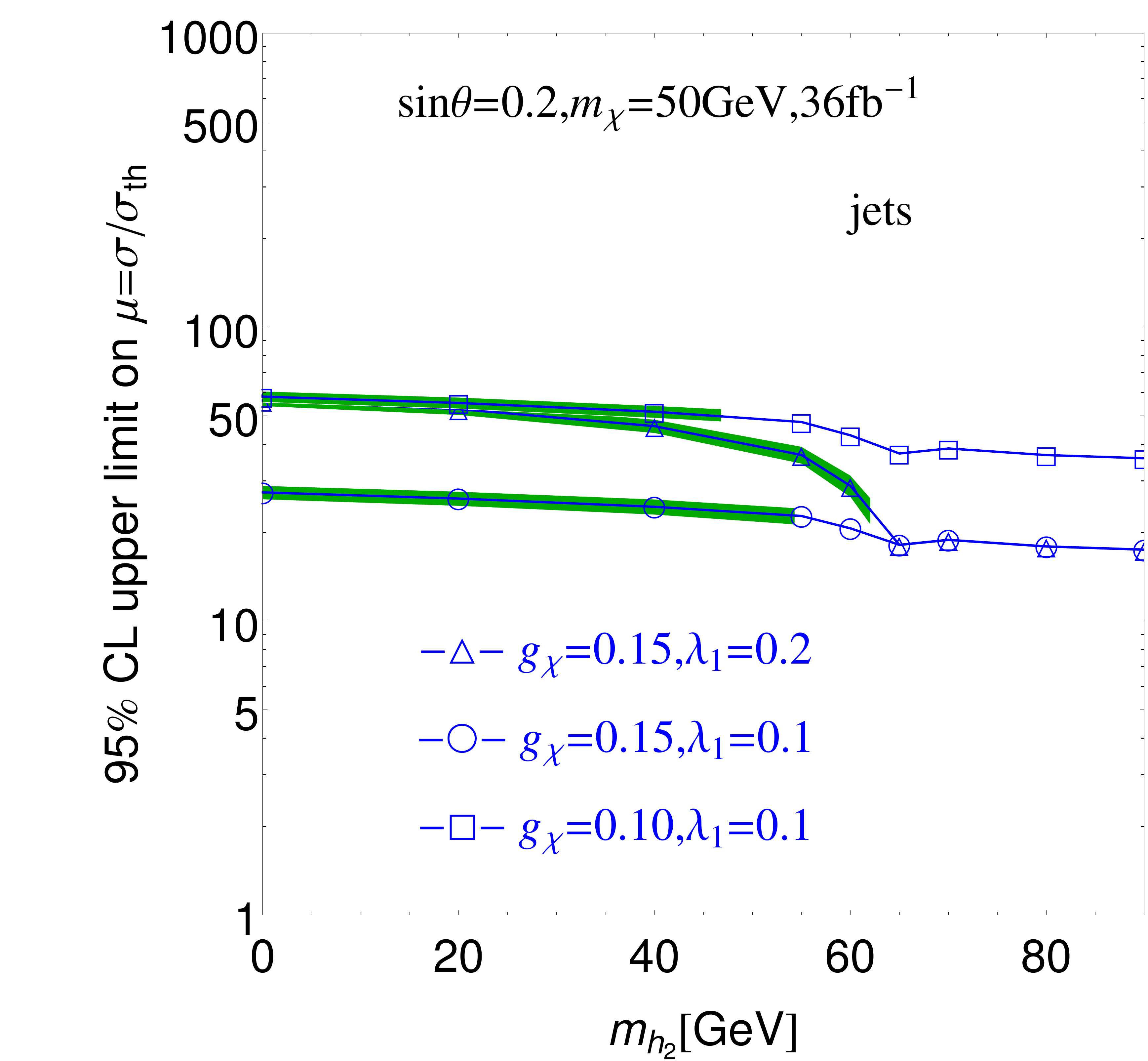}
\includegraphics[width=0.33\textwidth]{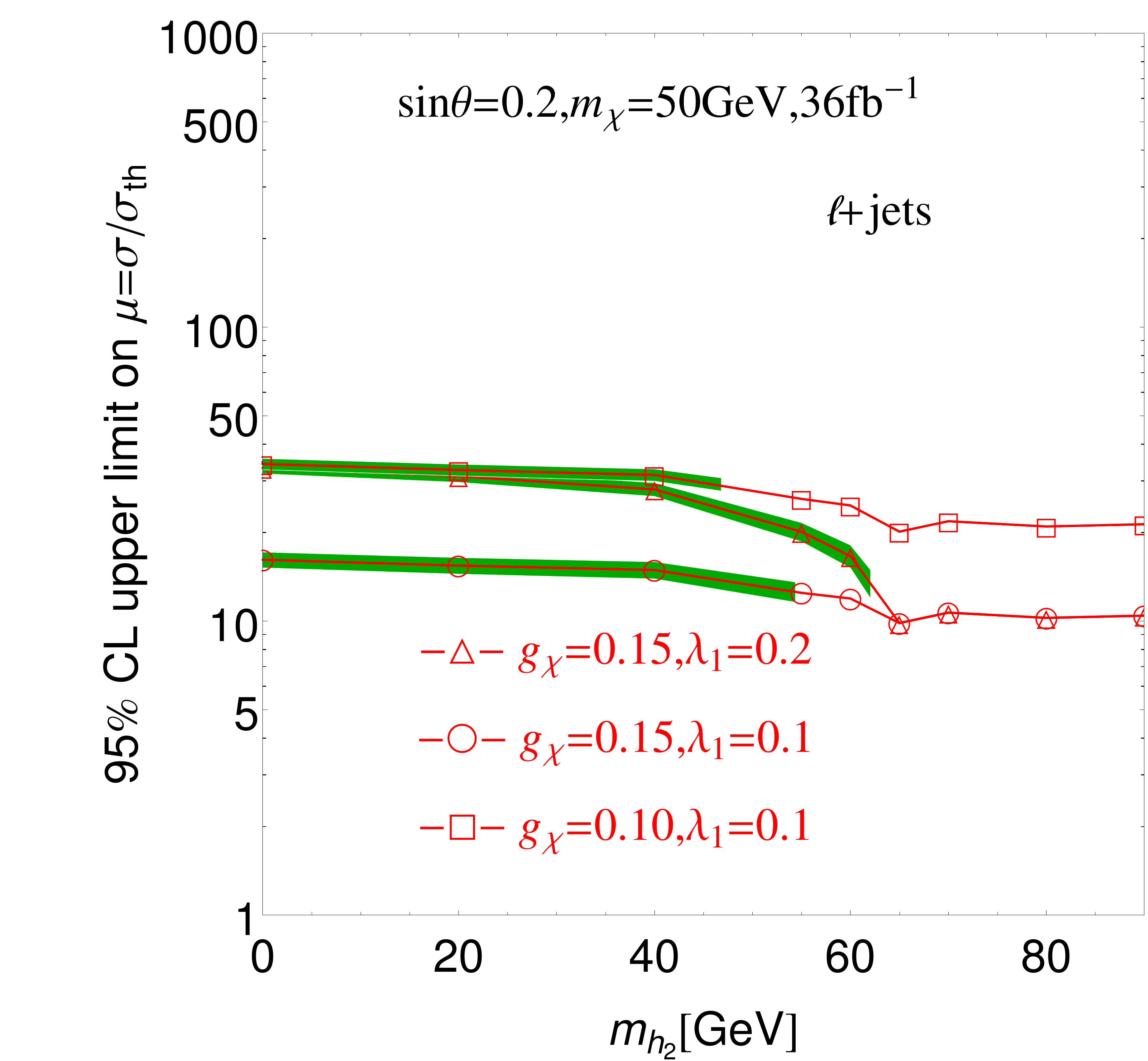}
\caption{The 95\% C.L. upper limits on $\mu$ in the inclusive hadronic (jets) and semileptonic
$(\ell+\text{jets})$ channels~\cite{Sirunyan:2017xgm,CMS:2016mxc}. Left panel: The upper limits for
Case A in the SFDM model (solid curves) and in the simplified model (dashed curves). Middle and
right panels: The upper limits on $\mu$ for Case B in the SFDM model for three benchmark values of
$g_{\chi}$ and $\lambda_1$. Parameter space of $m_{h_2}$ denoted as thick green curves is excluded by the measurements of Higgs BSM decay branching ratio.}
\label{fig:upper_limit2}
\end{figure}

In Fig.~\ref{fig:upper_limit2}, we show the 95\% C.L. upper limits on $\mu$ in the inclusive hadronic
(jets) and semileptonic $(\ell+\text{jets})$ channels~\cite{Sirunyan:2017xgm,CMS:2016mxc} in the
SFDM model with two mediators as well as one mediator $h_2$ (identified as the simplified model)
with the integrated luminosity of $36~\text{fb}^{-1}$ for Case A and Case B. In the left panel, we find
that, although the couplings of the mediator(s) to the SM quarks and DM are small for Case A,
the 95\% C.L. upper limit on $\mu$ in the SFDM model can reach $\mu<10$ and is nearly
independent of $m_{h_2}$, while the upper limit in the simplified model is much weaker especially
for a larger $m_{h_2}$.
In the middle and right panels, we display the upper limits of the SFDM model for three benchmark
values of $g_{\chi}$ and $\lambda_1$ for Case B. The measurements from the SM Higgs BSM
decay branching ratio ($\text{Br}_{h_1}^{\text{BSM}}<0.34$~\cite{Khachatryan:2016vau}) can also
constrain the parameter space for $m_{h_2}<m_{h_1}/2$ denoted as thick green curves.
For smaller $\lambda_2$, a more severe bound on $\mu$ can be obtained. The upper limit is sensitive to
$m_{h_2}$ only if $m_{h_2}<m_{h_1}/2$. Roughly speaking, the upper limit with the integrated
luminosity of $36~\text{fb}^{-1}$ is below 50 with these three benchmark values for Case B.

\begin{figure}[!thb]
\includegraphics[width=0.40\textwidth]{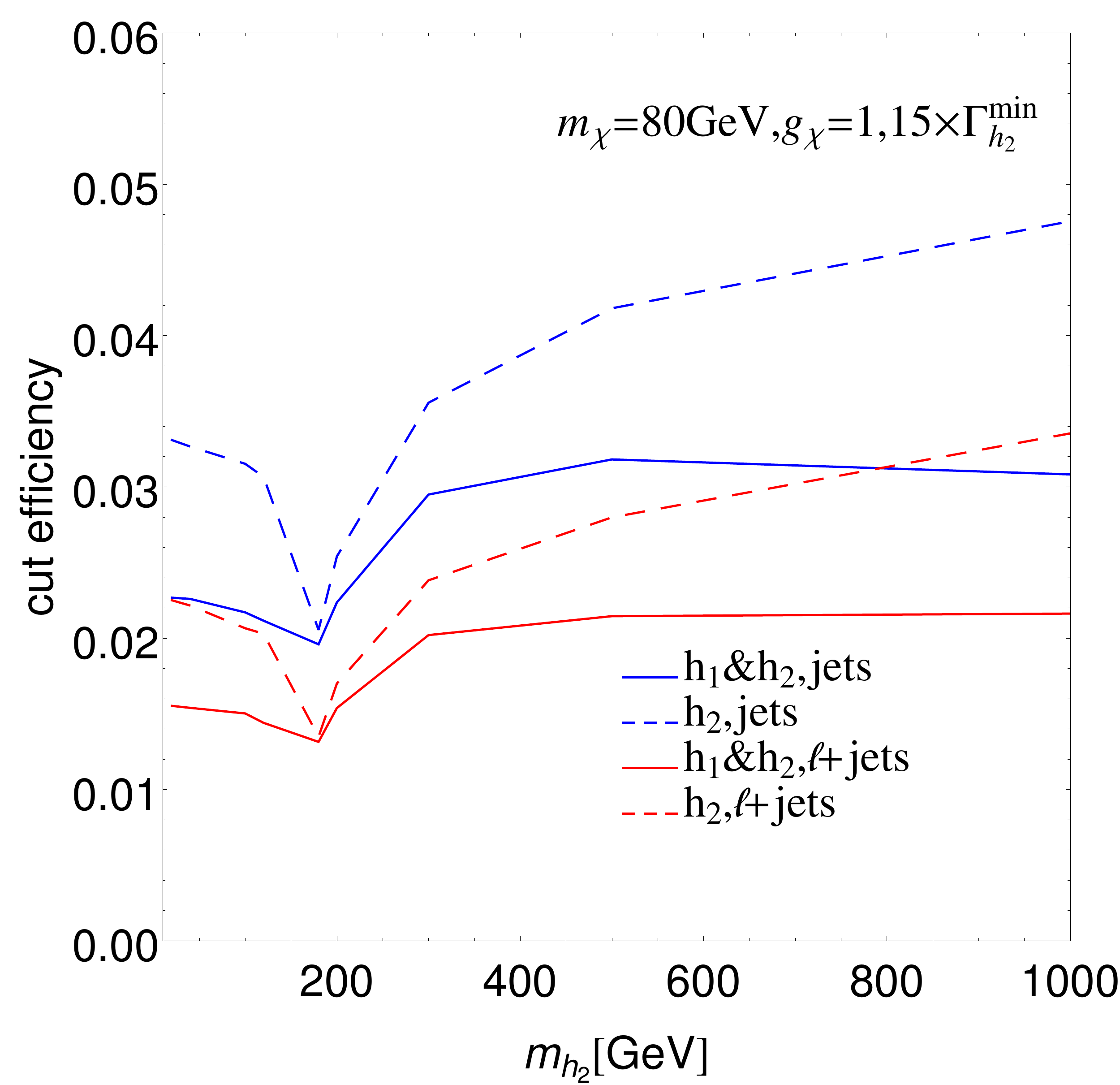}
\includegraphics[width=0.40\textwidth]{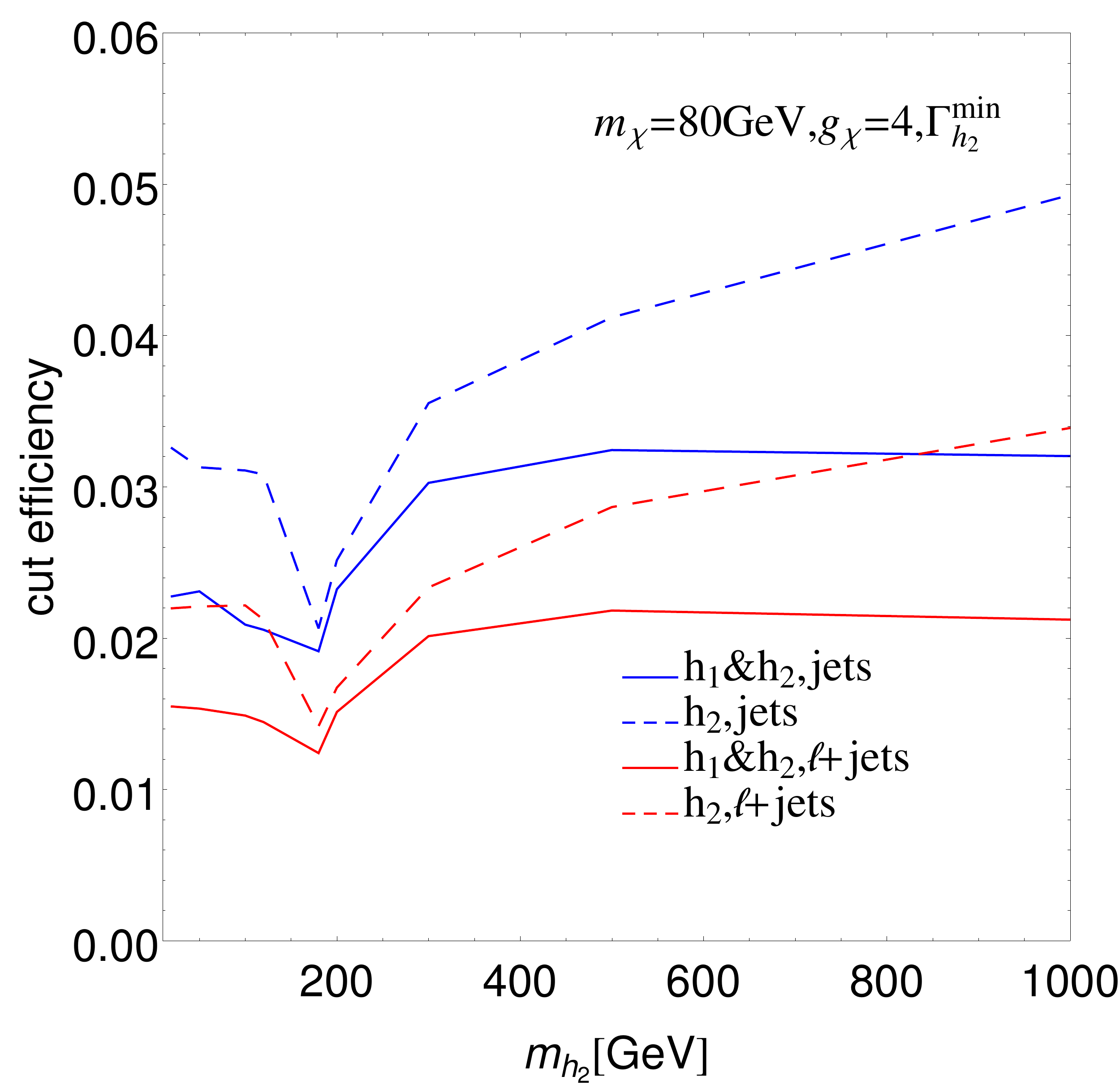}
\caption{Cut efficiencies for Case C and Case D in the SFDM model and simplified model in the
inclusive hadronic (jets) and semileptonic ($\ell$+jets) channels. }
\label{fig:cut_efficiency}
\end{figure}

\begin{figure}[!thb]
\includegraphics[width=0.40\textwidth]{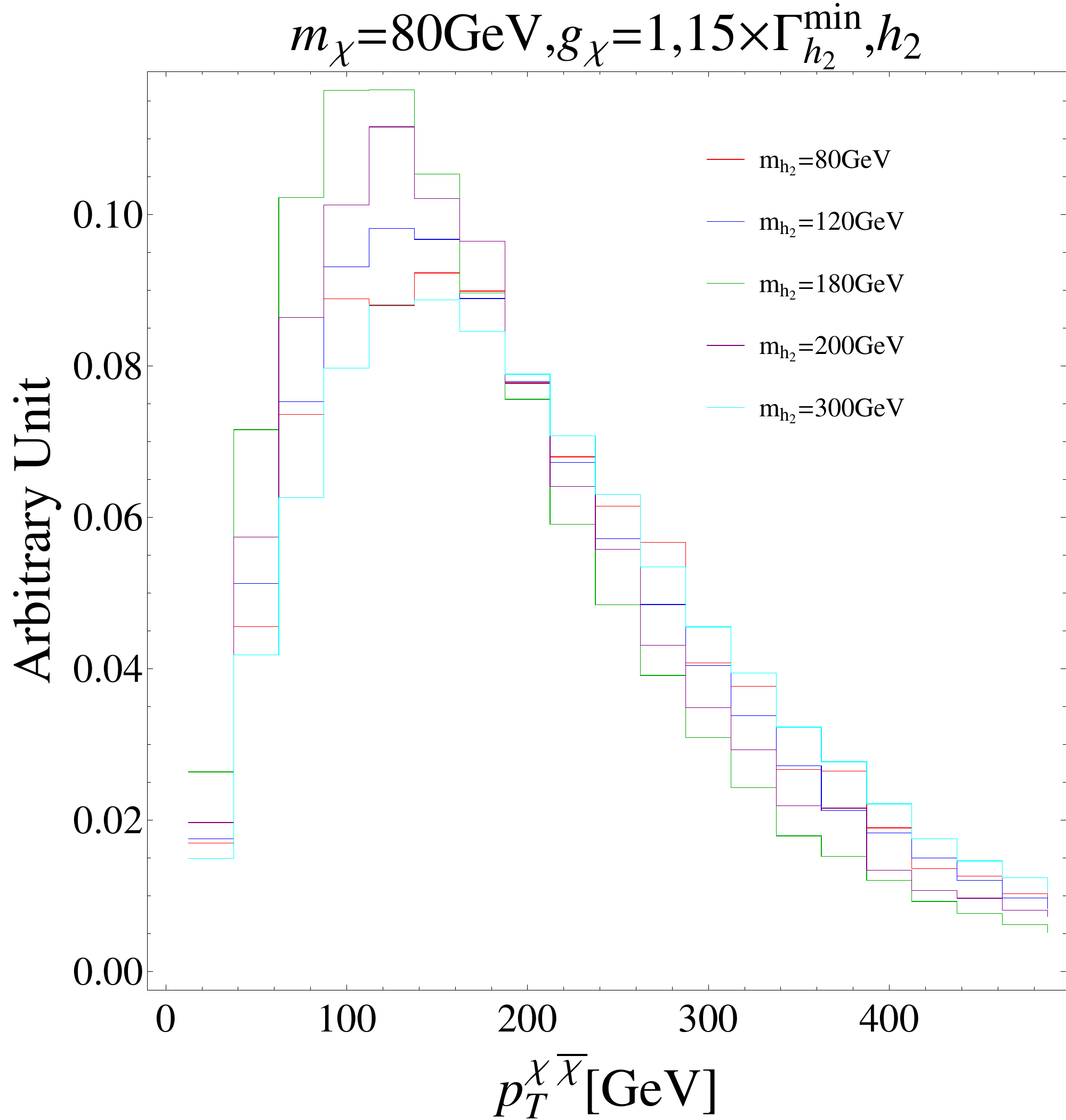}
\includegraphics[width=0.40\textwidth]{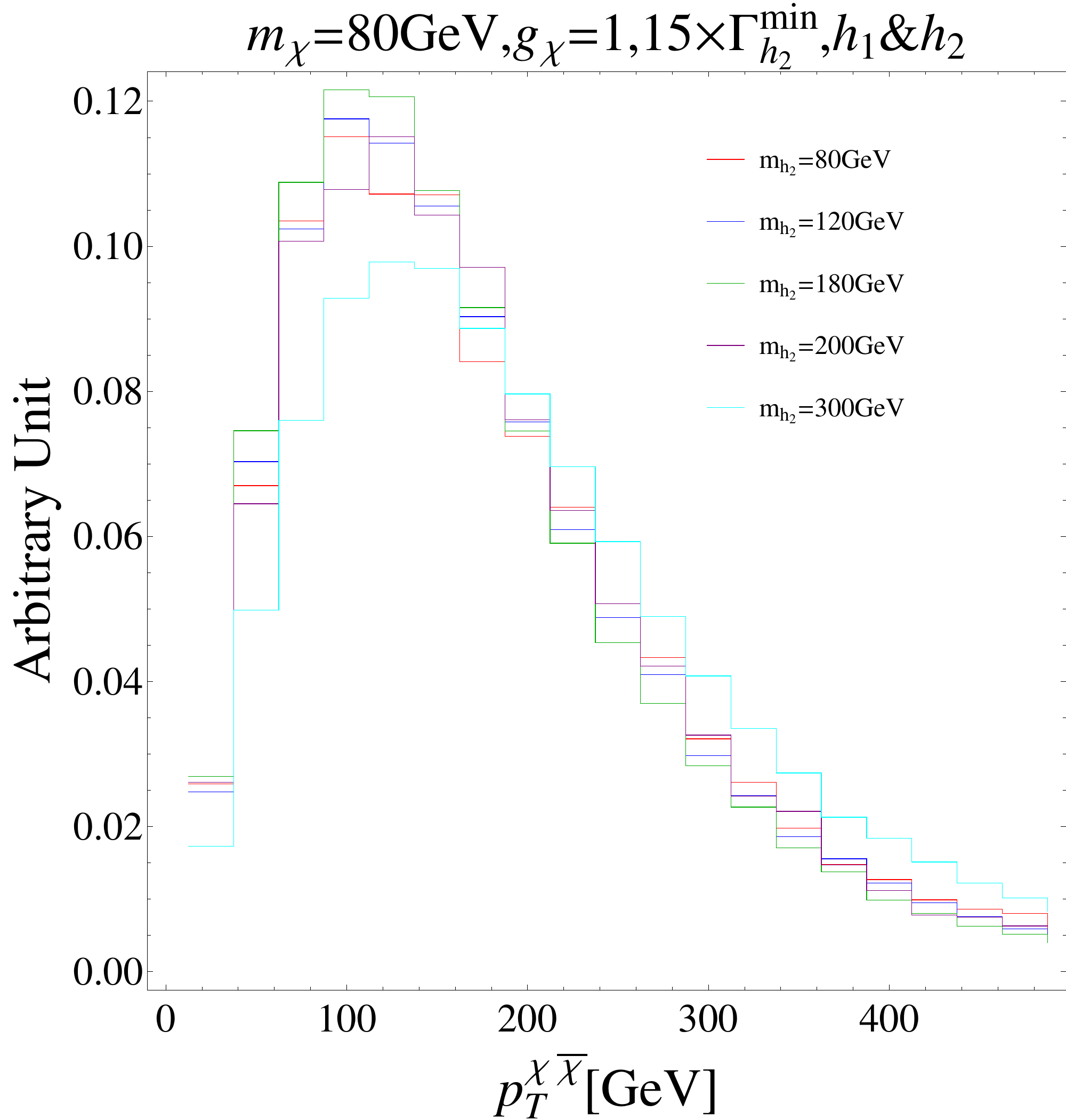}
\caption{The parton-level distributions of $p_{T}^{\chi\bar{\chi}}$ for Case C and Case D with
mediator $h_2$ (left) and two mediators $h_1$ and $h_2$ (right) for different $m_{h_2}$.}
\label{fig:ptn1n12}
\end{figure}

For Case C and Case D, both $h_1$ and $h_2$ play important roles in $pp\to t\bar{t}\chi\bar{\chi}$
in the SFDM model. Figure~\ref{fig:cut_efficiency} shows the cut efficiencies for Case C and Case D
in the inclusive hadronic and semileptonic channels in the SFDM model with two mediators as well as
one mediator $h_2$. Since the $p_{T}^{\chi\bar{\chi}}$ distribution
in the SFDM model is softer than that in the simplified model (see Fig.~\ref{fig:ptn1n1}), a lower cut
efficiency is achieved in the former scenario. On the other hand, in both scenarios the cut efficiencies
are lowest when $m_{h_2}$ is around $180\gev$. This is because the DM pairs $\chi\bar{\chi}$ are
mostly produced through the on-shell $h_2$ mediation while events with $m_{\chi\bar{\chi}} >
m_{h_2}$ are suppressed by the destructive interference between $h_1$ and $h_2$ in the SFDM
model and the $h_2$ propagator in the simplified model.  As a result, the $p_{T}^{\chi\bar{\chi}}$
distribution is softer for smaller $m_{h_2}$ for $m_{h_2}> 2 m_\chi$. On the other hand, when
$m_{h_2}< 2 m_\chi$, the DM pair can only be produced through off-shell $h_2$ mediation.
Then the relative suppression on event rates with higher $m_{\chi\bar{\chi}}$ is weaker for lighter
$h_2$, leading to harder $p_{T}^{\chi\bar{\chi}}$ spectra for lighter $h_2$.
This can be seen from Fig.~\ref{fig:ptn1n12}: For $m_{h_2}<180\gev$, the $p_{T}^{\chi\bar{\chi}}$
spectrum decreases with $m_{h_2}$. On the other hand, for $m_{h_2}>180\gev$,
$p_{T}^{\chi\bar{\chi}}$ gets increased for larger $m_{h_2}$.

\begin{figure}[!thb]
\includegraphics[width=0.40\textwidth]{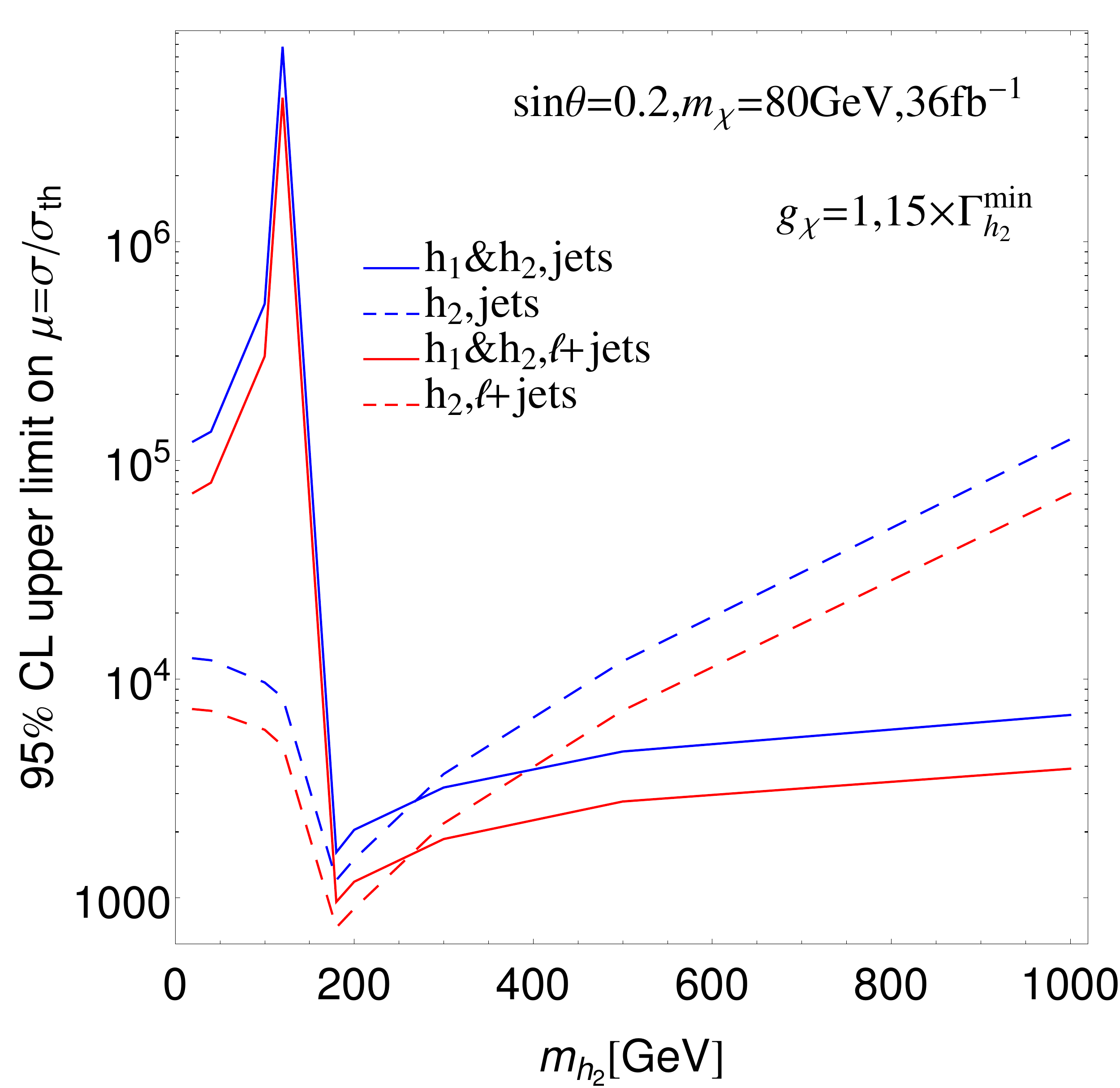}
\includegraphics[width=0.40\textwidth]{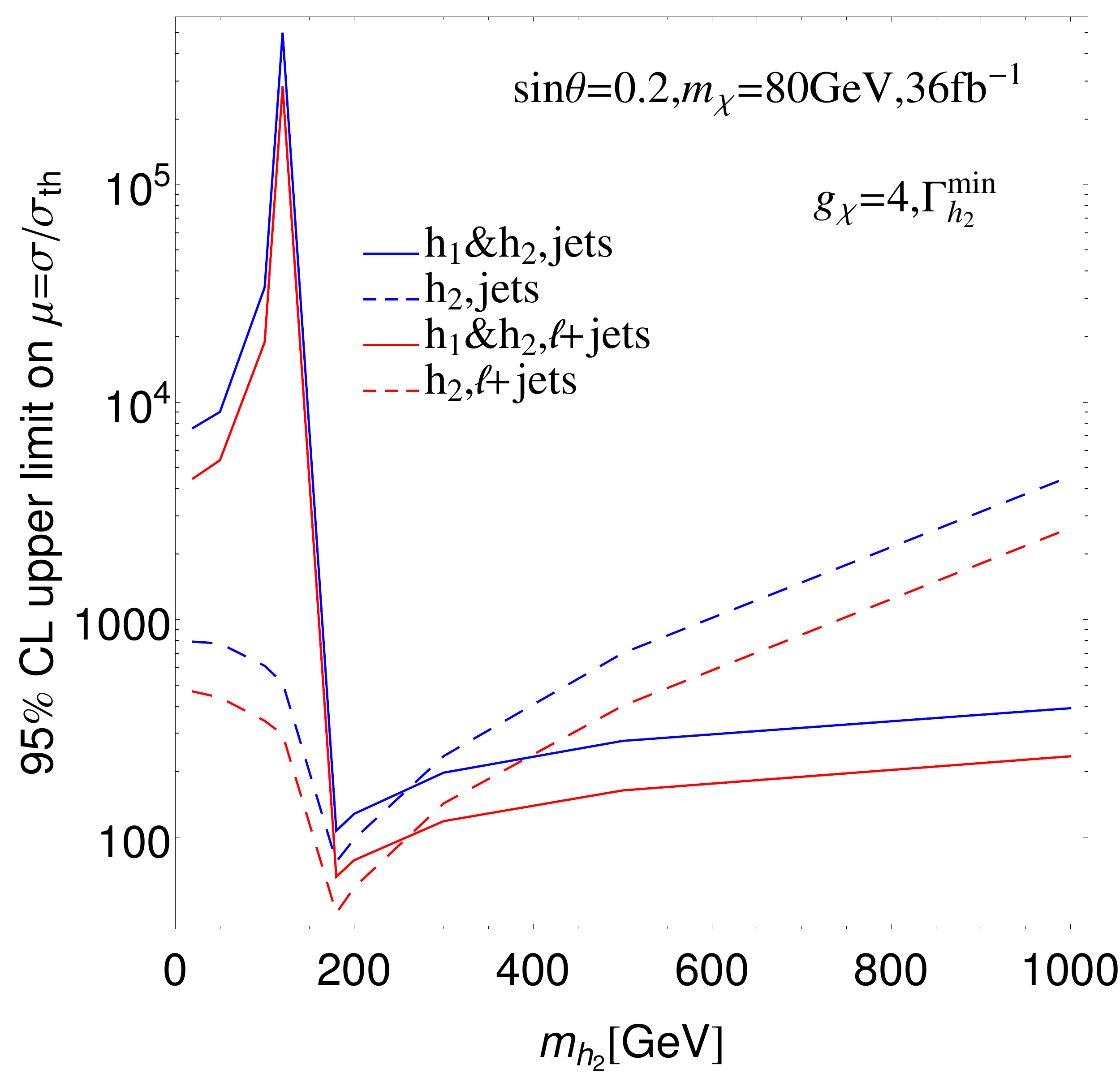}
\caption{The 95\% C.L. upper limits on $\mu$ in the inclusive hadronic (jets) and semileptonic
$(\ell+\text{jets})$ channels~\cite{Sirunyan:2017xgm,CMS:2016mxc} for Case C and Case D in the
SFDM model (solid curves) and simplified model (dashed curves).}
\label{fig:upper_limit}
\end{figure}

Finally, we show the upper limits on $\mu$ for Case C and Case D in the inclusive hadronic and
semileptonic channels with the integrated luminosity of $36\fbi$ in Fig.~\ref{fig:upper_limit}.
Due to the destructive interference between diagrams with $h_1$ and $h_2$ mediation, the LHC
search sensitivities on the SFDM model are extremely weak in the region of $m_{h_2} < 2 m_\chi$.
Without the destructive interference effects as in the simplified model, the sensitivities in the same
region can be more than order of magnitude better, but still way below the LHC probe at the current stage.
For $m_{h_2}\gtrsim 2m_{\chi}$, the interference effects on the total cross section can be destructive
or constructive in the SFDM model depending on $m_{h_2}$ as shown in the right panel of
Fig.~\ref{fig:ratio}. However, the interference effects always reduce the cut efficiency due to the
softened energy scale as compared to that in the simplified model. Both facts lead to a  better
sensitivity in the SFDM model than that in the simplified model for $m_{h_2}\gtrsim 300\gev$, and
becomes opposite for $m_{h_2}\lesssim 300\gev$.

\section{Summary}
\label{sec:summary}

In this work, we have studied the impact of the 125~GeV Higgs boson on searches for DM in association with a top pair (DM+$t\bar{t}$) at the LHC in the SFDM model with the Higgs portal.
Depending on the mass relations of two mediators and the DM, four cases are considered.
For Case A and Case B where the 125~GeV Higgs boson $h_1$ is on-shell, the DM production is
dominated by the mediator $h_1$. For Case C and Case D, $h_1$ is always off-shell while the
mediator $h_2$ can be either on-shell or off-shell. The impact of $h_1$ is significant in certain
parameter space in these cases, and the simplified model is not good enough.

Specifically, we find that when both $h_1$ and $h_2$ are off-shell (Case D), the destructive
interference makes the total cross section much smaller than that in the simplified model without
$h_2$. If only $h_2$ is on-shell (Case C),
the effect of $h_1$ on the total cross section becomes more important for larger $m_{h_2}$.
Besides, with a larger total width of $h_2$, which may come from a large coupling $g_{\chi}$ or
dominant decay of $h_2$ into the extra dark sector particles, the relative contribution of
$h_1$ ($h_2$) to the total cross section for Case C is further increased (decreased).
It is found that irrespective of $g_{\chi}$ and $\Gamma_{h_2}$ the interference effect for Case C is
destructive in the region of $2m_{\chi}\lesssim m_{h_2}\lesssim 380\gev $ and constructive for
$m_{h_2}\gtrsim 380\gev$ with $\sin\theta=0.2$ and $m_{\chi}=80\gev$. In addition to the total cross
section, $h_1$ can also affect the differential distribution of the DM+$t\bar{t}$ process.
Especially, the $p_{T}^{\chi\bar{\chi}}$ in the SFDM model is always soften as compared to that
in the simplified model for Case C and Case D.

Finally, we study the impact of $h_1$ on the LHC bounds of the DM+$t\bar{t}$ search in the inclusive
hadronic and semileptonic channels with the integrated luminosity of $36\fbi$. We find that the upper
limit on the signal strength $\mu$ for Case A in the SFDM model is smaller than 10, which is almost
independent of $m_{h_2}$. For Case B, the sensitivity also depends on the triple scalar coupling $\lambda_1$ of
$h_1-h_2-h_2$. Roughly, the upper limit is below 50 for the benchmark values discussed. For Case C, the sensitivity in the SFDM model
is extremely weak as compared to that in the simplified model due to the destructive interference
between the SM Higgs boson and the singlet scalar, which were largely ignored in theoretical and
experimental papers except in Refs.~\cite{Chpoi:2013wga,Baek:2015lna,Ko:2016ybp,
Ko:2016xwd,Baek:2017vzd,Kamon:2017yfx,Dutta:2017sod}.
For Case D, the upper limit in the SFDM model is better than that in the simplified model in the
region of $m_{h_2}\gtrsim 300\gev$ and becomes opposite for $m_{h_2}\lesssim 300\gev$.

Before closing, we would like to point out that the 125~GeV Higgs boson is also important for the VDM search at high-energy colliders.  If one generates the vector DM mass by a
dark Higgs mechanism, then there will be a mixing between the dark Higgs boson and the SM
Higgs boson~\cite{Baek:2012se}, resulting in two scalar propagators that can produce interesting 
interference~\cite{Baek:2015lna,Ko:2016ybp}.~\footnote{Note that there is no need to consider two scalar propagators in case of real singlet 
scalar DM (see, for example, Ref.~\cite{Ko:2016xwd}).} 
Then the amplitude for the VDM pair production
at high-energy colliders will take a form similar to Eq.~\eqref{eq:amplitude}. Effects of these two scalar propagators 
have been studied in the context of characterizing the mass and the spin of the Higgs 
portal scalar, fermion and vector DM at the ILC~\cite{Ko:2016xwd,Kamon:2017yfx} and 
at the LHC and 100~TeV $pp$ collider~\cite{Dutta:2017sod}.

In conclusion, we would like to emphasize that the contribution of the 125~GeV Higgs boson 
should  be properly included to interpret correctly the LHC dark  matter searches in case of 
the $s$-channel scalar mediators:  It is important not only for the gauge invariance  and renormalizability at the high-energy scale, but also for the quantitative difference of 
the upper limits and kinematic  distributions. 

\acknowledgments
We would like to thank Cheng-Wei Chiang, Jiayin Gu, Jusak Tandean and Yi-Lei Tang for valuable discussions. GL is grateful to the KIAS members for kind hospitality. This work is supported in part the MOST MOST106-2112-M-002-003-MY3 (GL), by National Research
Foundation of Korea (NRF) Research Grant NRF-2015R1A2A1A05001869 (PK, JL) and by the NRF
grant funded by the Korea government (MSIP) (No. 2009-0083526) through Korea Neutrino Research
Center at Seoul National University (PK).

\appendix
\section{Relic density and direct detection cross section}
\label{app:relic_density}
In this Appendix, we show the relic densities and spin-independent direct detection cross sections calculated using \texttt{micrOMEGAs}~\cite{Belanger:2008sj} for the benchmark points  of the cases categorized in Section~\ref{sec:xsection}.

Apart from the SM fermions or gauge bosons, the DM pair can also annihilate into scalar bosons if it is kinematically allowed. The couplings of $h_1-h_2-h_2$  and $h_2-h_1-h_1$ are defined in Eq.~\eqref{eq:lam1-lam2}, while the couplings of $h_1-h_1-h_1$  and $h_2-h_2-h_2$ are given by
\begin{align}
\lambda_{111}=\lambda_H v_H c_\theta^3-\mu_1/2 s_{\theta} c_{\theta}^2+\lambda_{HS} v_H s_{\theta}^2 c_{\theta}-1/6\mu_2 s_{\theta}^3,\\
\lambda_{222}=\lambda_H v_H s_\theta^3+\mu_1/2 c_{\theta} s_{\theta}^2+\lambda_{HS} v_H c_{\theta}^2 s_{\theta}+1/6\mu_2 c_{\theta}^3.
\end{align}
We take $\lambda_{HS}=0.02$ and $\mu_2=100\gev$ so that the triple scalar couplings satisfy the experimental measurements. For instance, if $m_{h_2}=80\gev$, $\lambda_{111}=0.94\lambda_{\text{SM}}$, $\lambda_{122}=-0.097\lambda_{\text{SM}}$, $\lambda_{211}=0.44\lambda_{\text{SM}}$ and $\lambda_{222}=0.52\lambda_{\text{SM}}$ with $\lambda_{\text{SM}}$ defined in Eq.~\eqref{eq:lam1_lam2}.

For Case C and Case D,  a large coupling $g_{\chi}$ is allowed since $m_{h_1}<2m_{\chi}$. 
As a result, the relic density of $\chi$ is small and can be below the measured DM relic density 
$(\Omega_0 h^2=0.120\pm 0.001)$~\cite{Aghanim:2018eyx}. In Tab.~\ref{tbl:case C} and Tab.~\ref{tbl:case D}, 
we show the relic densities and direct detection cross sections for the benchmark points with 
$s_{\theta}=0.2$, $g_{\chi}=4$, $m_{\chi}=80\gev$ and $m_{H_2}\in [70,500]\gev$.
In Case C, the DM is annihilated away through the $s$-channel $h_{1,2}$ mediation. Given a large $g_\chi=4$,
the relic densities of all the benchmark points are below the measured DM relic density and are independent of the triple scalar couplings. 
In Case D, the relic densities for $m_{h_2}=70$ and $90\gev$ are far below the measured DM relic density due to the annihilation of $\chi\bar{\chi}\to h_2h_2$. This channel is kinematically suppressed for $m_{h_2} \gtrsim 90$ GeV. Then, the DM can only annihilate through the $s$-channel $h_{1,2}$ mediation as in Case C.
For $m_{h_2}=110$ and $130\gev$, the relic densities becomes much larger because of the cancellation between the contributions from the mediators $h_1$ and $h_2$. 

For Case A and Case B, since $m_{h_1}>2m_{\chi}$ the coupling $g_{\chi}$ is severely constrained by the measurements of Higgs invisible decay branching ratio. In Tab.~\ref{tbl:case B1}, we show the relic densities for benchmark points in Case B with $g_{\chi}=0.15$ and $m_{\chi}=50\gev$ (the relic densities for Case A, which are larger, are not shown here). The relic densities of all benchmark points are larger than the measured DM relic density. As we explained in Section~\ref{sec:SFDM}, this can be weakened with the opening of new DM annihilation channels such as $\chi\bar{\chi}\to Z^\prime Z^\prime$ or coannihilation within a richer dark sector.

We can find that the DM-nucleon scattering cross section is well described by 
\begin{align}
\sigma^{\text{SI}}_{p} \propto (g_\chi  \sin(2\theta))^2 ~\left( \frac{1}{m_{h_1}^2} - \frac{1}{m_{h_2}^2} \right)^2.
\end{align}
Benchmark points in all case are challenged by current DM direct detections~\cite{Akerib:2016vxi,Aprile:2017iyp,Cui:2017nnn,Aprile:2018dbl} (for comparison, the $\sigma^{\text{SI}}_{p}$ of points with $\Omega h^2 < 0.120$ should be rescaled by a factor $\Omega h^2 / 0.120$). This indicates that there will be other DM annihilation mechanisms if our DM indeed comprises a component of a full DM sector. 

\begin{table}[t!hb]
\tabcolsep=6pt
\caption{Relic densities and spin-independent direct detection cross sections for Case C with $s_{\theta}=0.2$, $g_{\chi}=4$ and $m_{\chi}=80\gev$.}
\begin{tabular}{c|cccc}
\hline
\hline
$m_{h_2}$ [GeV]   & 200	&	300	&	400	&	500
\\ \hline
$\Omega h^2$   & $2.56\times 10^{-2}$	&	$7.03\times 10^{-2}$	& $8.51\times 10^{-2}$		&	$9.19\times 10^{-2}$
\\ \hline
$\sigma^{\text{SI}}_{p}$ [pb] & $1.30\times 10^{-7}$ & $2.38\times 10^{-7}$ & $2.84\times 10^{-7}$ & $3.07\times 10^{-7}$
\\ 
\hline\hline
\end{tabular}
\label{tbl:case C}
\end{table}

\begin{table}[t!hb]
\tabcolsep=6pt
\caption{Relic densities and spin-independent direct detection cross sections for Case D with $s_{\theta}=0.2$, $g_{\chi}=4$, $m_{\chi}=80\gev$, $\lambda_{HS}=0.02$ and $\mu_2=100\gev$.}
\begin{tabular}{c|ccccc}
\hline
\hline
$m_{h_2}$ [GeV]   & 70	&	90	&	110	&	130 & 150
\\ \hline
$\Omega h^2$  & $1.87\times 10^{-5}$	&	$8.04\times 10^{-3}$	& 1.15 &	3.72 & $3.72\times 10^{-2}$
\\ \hline
$\sigma^{\text{SI}}_{p}$ [pb] & $1.67\times 10^{-6}$ & $3.01\times 10^{-7}$ & $2.96\times 10^{-8}$ & $1.99\times 10^{-9}$ & $3.26\times 10^{-8}$
\\ 
\hline\hline
\end{tabular}
\label{tbl:case D}
\end{table}

\begin{table}[t!hb]
\tabcolsep=6pt
\caption{Relic densities and spin-independent direct detection cross sections for Case B with $s_{\theta}=0.2$, $g_{\chi}=0.15$, $m_{\chi}=50\gev$, $\lambda_{HS}=0.02$ and $\mu_2=100\gev$.}
\begin{tabular}{c|cccc}
\hline
\hline
$m_{h_2}$ [GeV]    & 60	&	70	&	80	&	90 
\\ \hline
$\Omega h^2$ &	8.27 & 8.31 & 8.17 & 7.77 
\\ \hline
$\sigma^{\text{SI}}_{p}$ [pb] & $5.40\times 10^{-9}$ & $2.32\times 10^{-9}$ & $1.01\times 10^{-9}$ & $4.18\times 10^{-10}$
\\ 
\hline\hline
\end{tabular}
\label{tbl:case B1}
\end{table}

\bibliographystyle{apsrev4-1}

\bibliography{reference}

\end{document}